\documentclass[manuscript,screen, nonacm]{acmart}

\AtBeginDocument{%
  \providecommand\BibTeX{{%
    \normalfont B\kern-0.5em{\scshape i\kern-0.25em b}\kern-0.8em\TeX}}}

\setcopyright{acmcopyright}
\copyrightyear{2023}
\acmYear{2023}
\acmDOI{XXXXXXX.XXXXXXX}

\usepackage{url}
\usepackage{graphicx}
\usepackage{amsmath}
\usepackage{amsfonts}
\usepackage{booktabs}
\usepackage{subfigure}
\usepackage{array}
\usepackage{multirow}
\usepackage{etex}
\usepackage{cleveref}

\newcolumntype{P}[1]{>{\centering\arraybackslash}p{#1}}
\usepackage{pgfplots}
\usepgfplotslibrary{dateplot}
\pgfplotsset{compat=1.16}
\usepackage{xcolor}
\usepackage{hyperref}

\usepackage{geometry}
\begin{document}

\title{AI-powered Fraud Detection in Decentralized Finance: A Project Life Cycle Perspective}

\author{Bingqiao Luo}
\email{luo.bingqiao@u.nus.edu}

\author{Zhen Zhang}
\email{zhen@nus.edu.sg}

\author{Qian Wang}
\email{qiansoc@nus.edu.sg}

\author{Anli Ke}
\email{Anli.ke@aalto.fi}

\author{Shengliang Lu}
\email{lusl@nus.edu.sg}

\author{Bingsheng He}
\email{hebs@comp.nus.edu.sg}

\affiliation{%
  \institution{National University of Singapore}
  \country{Singapore}}

\date{}
\begin{abstract}
In recent years, blockchain technology has introduced decentralized finance (DeFi) as an alternative to traditional financial systems. DeFi aims to create a transparent and efficient financial ecosystem using smart contracts and emerging decentralized applications. However, the growing popularity of DeFi has made it a target for fraudulent activities, resulting in losses of billions of dollars due to various types of frauds. To address these issues, researchers have explored the potential of artificial intelligence (AI) approaches to detect such fraudulent activities. Yet, there is a lack of a systematic survey to organize and summarize those existing works and to identify the future research opportunities. In this survey, we provide a systematic taxonomy of various frauds in the DeFi ecosystem, categorized by the different stages of a DeFi project's life cycle: project development, introduction, growth, maturity, and decline. This taxonomy is based on our finding: many frauds have strong correlations with the stages of the DeFi project. According to the taxonomy, we review existing AI-powered detection methods, including statistical modeling, natural language processing and other machine learning techniques, etc. We find that fraud detection in different stages employs distinct types of methods and observe the commendable performance of tree-based and graph-related models in tackling fraud detection tasks. By analyzing the challenges and trends, we present the findings to provide proactive suggestion and guide future research in DeFi fraud detection. We believe that this survey is able to support researchers, practitioners, and regulators in establishing a secure and trustworthy DeFi ecosystem.
\end{abstract}

\maketitle

\section{Introduction}
\label{sec:introduction}
\begin{figure}
\centering 
\setlength{\abovecaptionskip}{0cm}
\setlength{\belowcaptionskip}{0cm}
\includegraphics[scale=0.28]{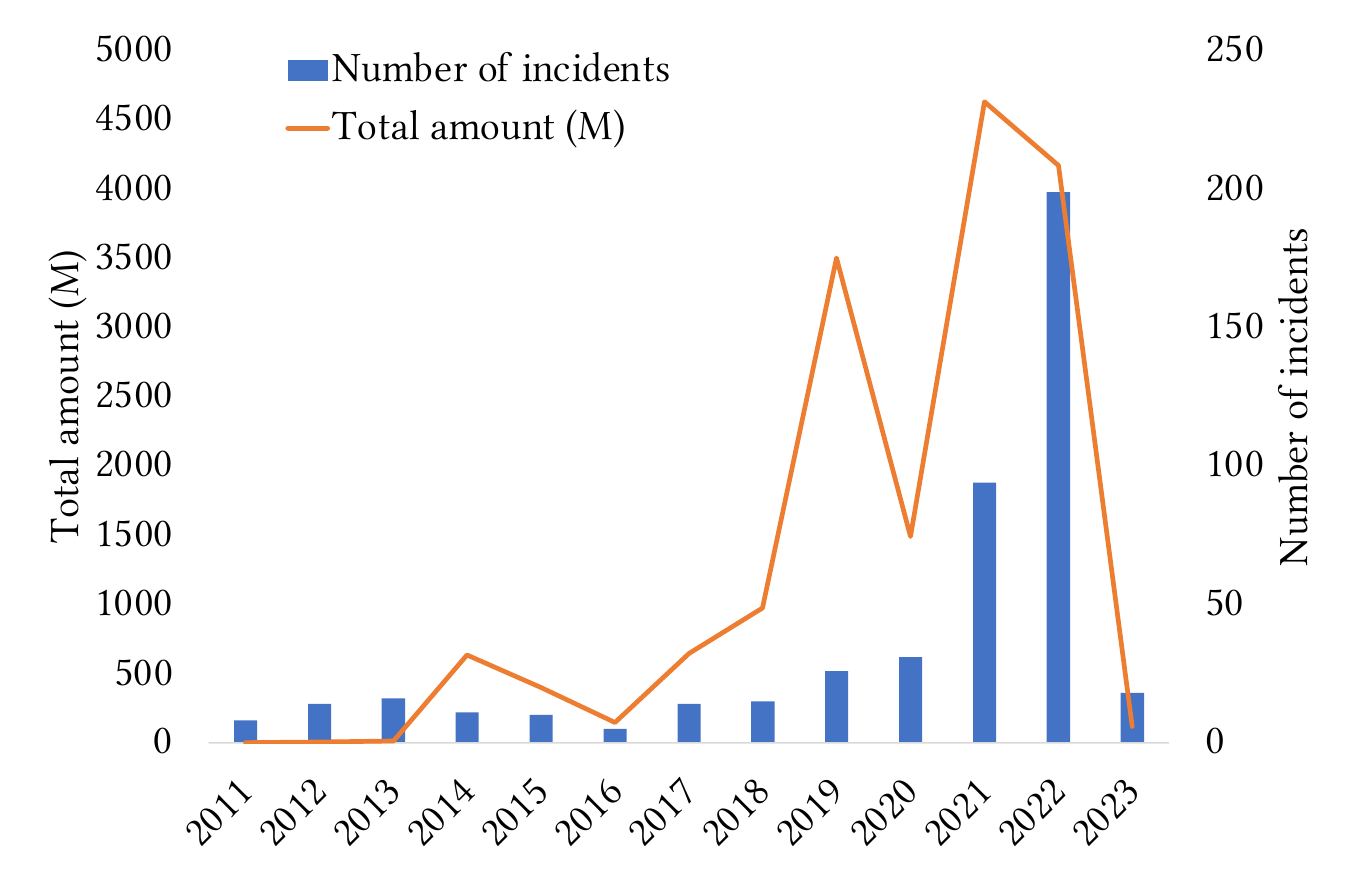} 
\caption{Number and Volume of DeFi Incidents (Data from \cite{crystal}, covering 21/06/2011 to 18/02/2023).} 
\label{fig:number} 
\vspace{-1em}
\end{figure}

With the rapid development of blockchain technology, decentralized finance (DeFi) has become an emerging alternative to address the challenges faced by traditional finance (CeFi) systems. While CeFi relies on intermediaries in a centralized system for transactions, DeFi uses distributed ledger technology, smart contracts, and a wide range of decentralized apps (DApps) to enable global, peer-to-peer financial activities, eliminating traditional intermediaries for transparency and efficiency. As reported by Dune \cite{dune}, the number of unique addresses in DeFi has risen to over 42 million as of July 2023, a significant increase from the 20 million addresses recorded at the start of 2022. Yet, with its increasing popularity, fraudulent activities in the DeFi ecosystem have surged, posing significant risks and eroding trust, impacting both individual users and the entire DeFi community. As shown in \autoref{fig:number}, from the first crypto exchange hack (Mt. Gox) in 2011 up until 2023 February, DeFi fraud has experienced a significant increase. Specifically, a total of 231 DeFi hacks, 135 security attacks, and 95 fraudulent schemes have been reported, resulting in approximate \$16.7 billion in stolen cryptocurrencies over the span of 12 years \cite{crystal}. This emphasizes the importance of studying these frauds and developing effective methods to detect them, in order to create a more DeFi trustworthy ecosystem.

With rising fraud activities, advancements in Artificial Intelligence (AI) have demonstrated great potential in tackling complex problems for fraud detection in recent years. Researchers have used these technologies to enhance fraud detection within the DeFi ecosystem using publicly available data. Machine learning models, such as tree-based and deep learning approaches, have been applied to classify suspicious transactions \cite{weber2019anti, lo2023inspection} and identify potential fraudsters \cite{farrugia2020detection, ibrahim2021illicit}. Meanwhile, Natural Language Processing (NLP) techniques have been employed to analyze textual information from project whitepapers \cite{bian2018icorating}, social media posts \cite{toma2020initial}, and the code of smart contracts \cite{chen2021sadponzi, chen2021improving}. Other technologies, such as statistical modeling \cite{cong2021crypto} and network analysis \cite{zwang2018detecting}, have also been utilized to identify frauds within DeFi. Hence, conducting a comprehensive survey is imperative to collect existing fraud detection methods, review their effectiveness, identify gaps, and provide insights into future research directions.

We aim to provide a comprehensive overview of the current landscape. To the best of our knowledge, there is no systematic discussion providing a comprehensive taxonomy of fraud, which is crucial for a deeper understanding of DeFi frauds across the entire ecosystem. Previous literature \cite{badawi2020cryptocurrencies, bartoletti2021cryptocurrency, eigelshoven2021cryptocurrency} has primarily focused on detecting frauds in cryptocurrency markets, overlooking the frauds that occur in other financial components of the DeFi ecosystem. For example, each Non-Fungible Token (NFT) item is unique and non-fungible, resulting in the absence of standardized pricing and the reshaping of market dynamics. Another example is Decentralized Exchange (DEX), which creates a less-regulated realm that is vulnerable to exploitation due to its decentralized, smart contract-driven framework. These distinct components foster new forms of fraud, such as rug pulls \cite{xia2021trade} and DEX hacks \cite{daian2020flash}, while traditional financial scams, such as phishing \cite{saha2023demystifying} and wash trading \cite{cong2021crypto}, take on new forms. Moreover, while some existing surveys \cite{badawi2020cryptocurrencies, trozze2022cryptocurrencies} have examined countermeasures for individual frauds, they lack a holistic understanding of the nature of these frauds and a comprehensive classification methodology. Therefore, on the one hand, various frauds are still evolving in many DeFi projects. On the other hand, we have also witnessed a lot of research efforts have been devoted to developing AI-powered fraud detection in DeFi projects. However, there is no systematic survey to summarize those efforts and to identify the future research opportunities. This motivates us to comprehensively survey AI-powered approaches for DeFi fraud detection.

In this survey, we present a novel methodology for reviewing financial frauds across five stages in the life cycle of DeFi projects, based on the Product Life Cycle Theory \cite{SurveyMonkey}: development, introduction, growth, maturity, and decline. This taxonomy is based on our finding that many frauds have strong correlations with the DeFi project stage. Specifically, (1) in the project development stage, scammers exploit the lack of oversight and security measures to create investment traps, leading to fraudulent activities like Ponzi schemes \cite{vasek2015there} and rug pulls \cite{xia2021trade}. (2) When the project progresses through the introduction and growth stage, dishonest participants can manipulate the market using tactics such as wash trading \cite{aloosh2019direct} and insider trading \cite{felez2022insider} due to the less-regulated environment. (3) Moving into the maturity and decline stage, low-cap projects become susceptible to pump and dump schemes \cite{xu2019anatomy} due to relatively easier market conditions, while high-cap projects face risks such as money laundering \cite{weber2019anti} and phishing scams \cite{chen2020phishing}, which demonstrated new forms and characteristics in DeFi. This taxonomy provides a systematic categorization of DeFi frauds based on their occurrence time, revealing the complex landscape of DeFi frauds.

With this taxonomy, our survey provides a comprehensive review of on-chain frauds and existing fraud detection methods targeted at frauds that occur at each stage, leading to the discovery of noteworthy findings and potential future directions across the life cycle of DeFi projects. Below are some highlights: 
\begin{itemize}
    \item \textbf{Fraud Types and Organizers}: During the development stage, the majority of frauds involve investment scams organized by project developers. As projects progress to the introduction and growth stages, market manipulations emerge. Project developers and DeFi platforms collaborate during this phase to organize fraudulent activities. In the maturity and decline stages, the role of fraud organizers shifts to external entities, who also collaborate with DeFi platforms to organize some market manipulations and new forms of illicit activities.
    
    \item \textbf{Detection Methods}: During the development stage, tree-based models are prominent for fraud detection tasks because they have demonstrated superior performance for small data in this stage. In the introduction and growth stages, statistical methods and graph-based patterns are widely used to capture suspicious market activities. Advanced techniques, particularly graph-based methods, gain prominence in the maturity and decline stages to capture complex relationships in large and complex data.
    
    \item \textbf{Detection Effectiveness}: For the frauds occurred in the early stages, fraud detection tasks are still in the beginning phases. This could be due to a limitation of transaction data as well as a scarcity of reliable ground truths. In comparison, frauds in the later stages have received significant attention, as well as good detection performance using a large amount of data and advanced models.

    \item \textbf{Future Directions}: During the early stages, leveraging the potential of pre-trained models and conducting early-stage detection, can aid in proactive fraud prevention. In later phase, researchers can leverage the power of large language models (LLM) or foundational models to gather more comprehensive information and develop advanced detection techniques. More future directions and open issues throughout the entire life cycle are also presented in this paper.
    
\end{itemize} 

The remainder of this paper is organized as follows. We provide a background on DeFi frauds and our motivation of this survey in \Cref{sec:background} and \Cref{sec:motivation}, respectively. \Cref{sec:overview} presents a DeFi fraud taxonomy and detection methods overview. \Cref{sec:development,sec:introduction_growth,sec:maturity_decline} review stage-specific frauds and detection methods. \Cref{sec:discussion} presents discussion and findings, and \Cref{sec:future} discusses potential future directions. Finally, \Cref{sec:conclusion} concludes this paper.

\section{Background}
\label{sec:background}

\subsection{Decentralized Finance and Cryptocurrency}
\label{sec:defi_crypto}
This subsection introduces and compares DeFi and cryptocurrency. DeFi, a new financial ecosystem, is based on secure distributed ledgers powered by blockchain technology. With a wide range of decentralized applications (dApps), it provides users with efficient transactions and enhanced control over their financial assets. Various cryptocurrencies, such as Ethereum and Solana, are used for payment and collateral in many DeFi projects. While it is based on cryptocurrency, DeFi has significant similarities and differences. The use of blockchain technology, digital existence, support for peer-to-peer transactions, and independence from intermediaries such as banks and other financial institutions are all similarities between these two. However, while cryptocurrency serves as a means of transaction and a preservation for value in the ecosystem, DeFi goes beyond by offering a wide range of financial products and services.

Here are some representative components and financial services that distinguish DeFi from general cryptocurrency:
\begin{itemize}
\item{Smart contracts}: Smart contracts are blockchain programs designed for specific functions. They are used in both cryptocurrency and DeFi, but in different ways. In cryptocurrency, smart contracts are used to automate transactions and support specialized wallets, such as multisignature wallets. While in DeFi, smart contracts are used as the system's foundation for a broader range of financial services such as lending and borrowing, liquidity provision, and the creation of a diverse range of products such as multiple NFT collections. Ethereum, the primary blockchain serving as the foundation for numerous DeFi projects, had over 44 million contracts deployed on it as of August 2022\footnote{\url{https://crypto.news/over-44-million-contracts-deployed-on-ethereum-since-launch/}. Accessed on Aug 16, 2023.}. Smart contracts are also supported by a few other blockchains, including Binance Smart Chain (BSC) and Solana. 

\item{Middleware}: Middleware solutions such as oracles and bridges are vital in the DeFi ecosystem to enable communication between different blockchain networks, applications, and the real world. Oracles, like Chainlink\footnote{\url{https://chain.link/}. Accessed on Aug 28, 2023.}, connect blockchains with the external world, while bridges connect between different blockchains. For instance, Wrapped Bitcoin (WBTC)\footnote{\url{https://wbtc.network/}. Assessed on Aug 29, 2023.} transforms BTC into an ERC-20 token that compatible with Ethereum, maintaining its value equivalence.

\item{Decentralized financial platforms}: Multiple decentralized platforms play an important role in DeFi. They support a variety of financial services, including trading, lending, and borrowing. All these services are automated through the use of smart contracts. A notable service is yield farming \cite{augustin2022yield} that enables users to earn rewards by depositing their digital assets. Additionally, some DeFi platforms, such as Aave\footnote{\url{https://aave.com/}. Accessed on Aug 15, 2023.}, support flash loans \cite{wang2021towards}, which are non-collateral loans that must be repaid within the same transaction block. It is frequently used for arbitrage, liquidation, or other short-term profit opportunities.

\item{Decentralized exchanges (DEXs)}: DEXs redefine traditional exchanges in DeFi. In contrast to centralized exchanges (CEXs), which can also facilitate cryptocurrency transactions but are controlled by a central authority, DEXs offer more efficient user experiences by eliminating the need for intermediaries through smart contracts. Moreover, unlike many individual transfers among users in cryptocurrency, the diverse functions in DeFi, such as the auction of NFTs and token swapping, emphasize the necessity of DEXs. A notable service of DEX is staking \cite{cong2022staking} that locks crypto assets to smart contracts of exchanges, in return for rewards. Uniswap, with over 5 million users by July 2023, stands out as a significant DEX\footnote{\url{https://dune.com/queries/1688660}. Accessed on Aug 16, 2023.}.

\item{Non-fungible tokens (NFTs)}: NFTs, being non-fungible digital assets, signify ownership of unique items. There are many distinct items in each NFT collection. Different from cryptocurrency, they are non-fungible which means there are no identical NFT items. Again, smart contracts support all NFT-related activities, including minting, buying, selling, and transferring. Some marketplaces support fractional NFT, which allows users to trade fractional ownership of the NFT. Bored Ape Yacht Club (BAYC)\footnote{\url{https://boredapeyachtclub.com/}. Accessed on Aug 16, 2023.} is a well-known NFT collection, with an average sale price of more than \$200K and owned by many famous celebrities. In 2021, NFTs grew rapidly, and by February 2022, the NFT market had outperformed the crypto market in terms of year-to-date returns \cite{nansen}.

\end{itemize}

In addition to these representative examples, the DeFi ecosystem encompasses other financial services, such as insurance services and derivatives. These financial products and services combine to form a diverse ecosystem with significant potential, resulting in several distinct factors that contribute to DeFi's rapid growth: First, DeFi uses smart contracts to provide their users with efficient, transparent, and diverse financial solutions. Second, the significant attention that DeFi receives from both the cryptocurrency community and celebrities attracts a large number of participants. Third, DeFi broadens the scope of investment opportunities by allowing users to trade fractional ownership and traditionally non-liquid assets by tokenizing real-world assets. However, as the DeFi ecosystem grows in popularity, it also gives rise to potential vulnerabilities and frauds.

\subsection{Big Events of Frauds along the Time of DeFi}
Despite its promising potential, the DeFi ecosystem's extensive functionality and diverse components pose a number of challenges. This section provides an overview of the emergence of DeFi frauds throughout DeFi's history in order to provide a deeper understanding of how these incidents occurred and their consequences.

The following observations about fraudulent activities in the DeFi ecosystem have been highlighted \cite{crystal}. First, as illustrated in \autoref{fig:number}, overall fraud activity in the DeFi market has increased significantly in recent years. Since 2011, security breaches have led to over \$4.5 billion in thefts, scams have caused over \$7.5 billion in losses, and DeFi hacks have resulted in over \$4.81 billion stolen. Second, there has been a surge in fraud within DeFi projects. Notably, after 2021, the preferred method of cryptocurrency theft shifted from infiltrating crypto-exchange security systems to DeFi hacks. In 2022, the proportion of CEX to DEX hacks stood at 1:3 highlighting the prevalence of DeFi hacks which continuously rise. Third, rug pulls \cite{Chainalysis2022}, which refer to the sudden disappearance or abandonment of DeFi projects, became increasingly popular among fraudsters during the first half of 2022, with Tornado Cash being the most widely used service for laundering funds. Moreover, NFTs experienced a significant trend in 2021, with a 1,785\% increase in market capitalization during the year. However, this popularity led to NFTs becoming a prime target for rug pull scammers, with a large number of NFTs becoming rug pulls.

\autoref{fig:hacks} illustrates significant DeFi events from the past 5 years to demonstrate the impact and evolution of major DeFi frauds throughout history, each of which cost millions of dollars \cite{frontal}. Three examples are detailed below: 

\begin{figure}
\centering 
\setlength{\belowcaptionskip}{0cm}
\includegraphics[scale=0.3]{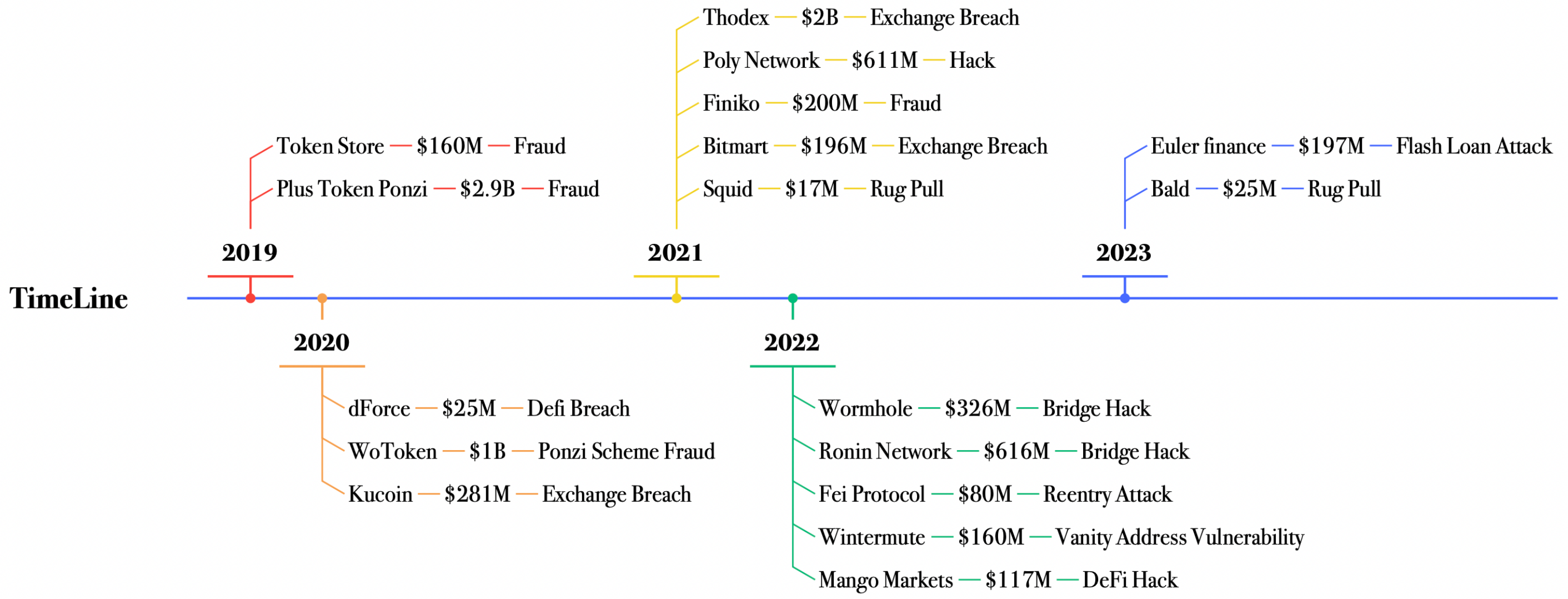} 
\caption{Big Events of Frauds in Recent 5 Years (Data from \cite{frontal}).} 
\label{fig:hacks} 
\vspace{-1.5em}
\end{figure}

\begin{itemize}
    \item Poly Network: Poly Network, a cross-chain protocol, experienced the largest hack in crypto history in August 2021. The hacker exploited a vulnerability in the protocol's "contract calls" and successfully stole \$611 million worth of various cryptocurrencies. However, the situation took an unexpected turn, as Poly Network's outreach efforts led to the hacker returning the stolen funds just three days later.
    
    \item Wormhole: Wormhole, a popular cross-chain bridge, experienced a significant incident in January 2022, causing a loss of \$326 million in Wrapped Ethereum (WETH). The hacker targeted the specific leg of Wormhole on the Solana blockchain, managing to mint WETH without locking up ETH in Wormhole. Jump Crypto, a stakeholder in Wormhole's development, replaced the stolen funds following the hack.
    
    \item Ronin: The Ronin sidechain, used for the play-to-earn game Axie Infinity, faced a major setback in March 2022 when it fell victim to a substantial hack. The attacker drained through "hacked private keys" and sign transactions from several validator nodes, resulting in a total loss of \$616 million. 
\end{itemize}

In summary, the growth of DeFi projects has attracted the attention of fraudsters, resulting in huge financial losses and serious trust concerns in the recent years. Understanding and detecting these frauds are critical for creating a more secure and trustworthy DeFi ecosystem.

\subsection{Factors Driving the Rise of Fraud in DeFi}
\label{sec:factors}
Given the prevalence of numerous frauds in the DeFi ecosystem, it naturally raises the question of why these incidents occur. In this subsection, we explore the factors driving DeFi frauds.  

We begin by examining the characteristics of DeFi compared to CeFi to investigate potential factors. While both CeFi and DeFi offer similar financial products and services to users, such as payment gateways, trading exchanges, and asset management tools, DeFi distinguishes itself from CeFi through its cryptocurrency-based nature and on-chain applications. A detail comparisons are illustrated in \autoref{fig:factors}. While some characteristics offer benefits to DeFi, they also introduce certain factors that drive fraudulent activities. Furthermore, as DeFi is a relatively new ecosystem, it encounters additional challenges and vulnerability. The combination of these factors has given rise to fraudulent activities within the DeFi space, as summarized below:

\begin{table}
\setlength{\abovecaptionskip}{0cm}
\setlength{\belowcaptionskip}{0cm}
\centering 
\includegraphics[scale=0.55]{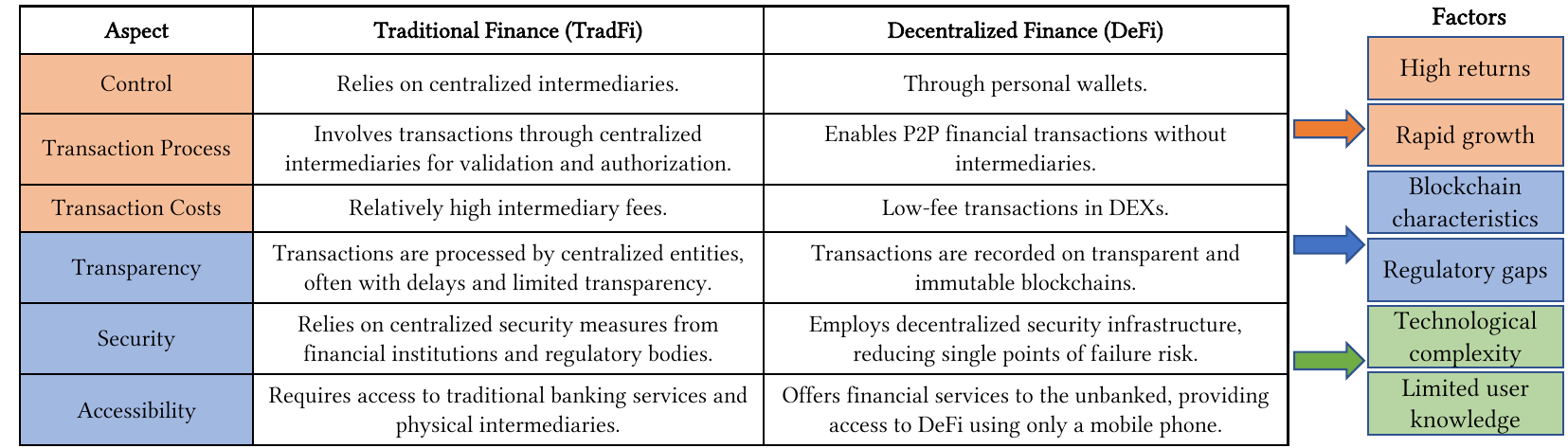} 
\caption{Comparisons of CeFi and DeFi and DeFi Fraud Factors} 
\label{fig:factors} 
\vspace{-1em}
\end{table}

\begin{enumerate}
    \item \textbf{High returns and rapid growth}: DeFi has seen rapid growth due to its easy control of financial assets, simple transaction process, and low transaction costs, with emerging new projects offering enticing investment opportunities. This growth, however, has attracted scammers who exploit unsuspecting investors. Both legitimate users and malicious actors are drawn to the allure of high returns and the expanding ecosystem.
    \item \textbf{Unique characteristics of Blockchain and regulatory gaps}: Participants in decentralized systems benefit from transparency, accessibility, anonymity, decentralization, and immutability. However, these features allow scammers and criminals to take advantage of the system by leveraging its decentralized nature to conduct fraudulent activities while concealing their true identity. Furthermore, because of blockchain's anonymity and decentralization, there is regulatory ambiguity for DeFi, allowing bad actors to operate in the shadows.
    \item \textbf{Technological complexity and limited user knowledge}: Moreover, many DeFi platforms use advanced techniques such as smart contracts, which can be complex and difficult for users to understand. Given that new DeFi users may be unfamiliar with the underlying technology, fraudsters have extensive knowledge in it and use it to deceive them, exacerbating the situation.
\end{enumerate}
In summary,  the interaction of high returns and rapid growth, the unique characteristics of blockchain, and technological complexity all contribute to DeFi frauds. These vulnerability highlight the importance of security measures and fraud detection strategies in the DeFi landscape.

\section{Motivation}
\label{sec:motivation}
As shown in \Cref{sec:background}, various types of frauds are emerging in DeFi, resulting in huge financial losses. Meanwhile, we've seen various AI methods for detecting and preventing these frauds have been developed. As a result, we are motivated to create a comprehensive survey to examine DeFi frauds and the corresponding AI detection methods.

We first collect and analyze related surveys in three areas: blockchain security, general cryptocurrency fraud, and DeFi financial fraud, in order to understand the current research landscape and identify current limitations. 

First, various surveys have explored security weaknesses in blockchain, including vulnerability assessments of blockchain elements \cite{hassan2022anomaly, leng2020blockchain, saad2019exploring}, attacks on specific blockchains \cite{conti2018survey, chen2020survey}, and security aspects related to cryptocurrency wallets \cite{Houy_Schmid_Bartel_2023} and smart contracts \cite{atzei2017survey, rouhani2019security}. However, these surveys do not focus on financial frauds within the DeFi space, which constitutes a unique aspect of our research objectives of surveying the financial frauds.

Second, scholars have concentrated on general cryptocurrency markets, investigating financial frauds and their detection methods, and categorizing by frauds or detection models. In terms of frauds, Badawi and Jourdan \cite{badawi2020cryptocurrencies} examined different cryptocurrency threats, encompassing types, scales, detection mechanisms, effectiveness, and public datasets. Trozze et al. \cite{trozze2022cryptocurrencies} categorized 29 fraud types via a survey of academic research, reports, publications, and alerts. Yan et al. \cite{yan2022blockchain} summarized abnormal behavior awareness methods and ideas on public and consortium blockchains, along with existing datasets related to mainstream blockchain security. Bartoletti et al. \cite{bartoletti2021cryptocurrency} devised an auto-scam detection tool using a scam report-based taxonomy, while Eigelshoven et al. \cite{eigelshoven2021cryptocurrency} explored cryptocurrency market manipulation. Krishnan et al. \cite{krishnan2023scams} analyzed 10 fraudulent activities and assessed machine learning detection models, though their performance comparison's value is limited due to different datasets. In terms of detection models, Li et al. \cite{li2020survey} categorized methods into universal and specific fraud detection techniques. Similarly, Li \cite{li2022survey} approached Ethereum's illicit detection from two angles: transaction data using general approaches and selected illicit transaction types.

Third, a recent study by Wu et al. \cite{wu2023financial} has presented a comprehensive taxonomy of security risks and financial crimes within the Web3 metaverse, encompassing DeFi as one of its applications. Their study addressed digital asset, smart contract, and blockchain frauds while offering insights on anti-crime measures from academic and government policy perspectives. Notably, their survey did not specifically assess AI-powered techniques targeting DeFi frauds, which distinguished our work in terms of motive. 

\begin{table}[h]
\setlength{\abovecaptionskip}{0cm}
\setlength{\belowcaptionskip}{-0.2cm}
\caption{Comparison of Existing Surveys on Cryptocurrency and DeFi Fraud. It covers target platforms, classification method, taxonomy through project life cycle, detail on-chain pattern exploration, AI detection models analysis, DeFi vs. CeFi fraud comparison, and discussion of challenges and future opportunities for each paper. For those classifying by detection methods, fraud taxonomy and on-chain patterns are not applicable (N/A) to them.}
\centering

\resizebox{\textwidth}{!}{%
\begin{tabular}{|m{4cm}<{\centering}|m{3.3cm}<{\centering}|m{3cm}<{\centering}|m{1.3cm}<{\centering}|m{1.4cm}<{\centering}|m{1.3cm}<{\centering}|m{1.4cm}<{\centering}|m{1.7cm}<{\centering}|}
\hline
\textbf{Paper} & \textbf{Platforms} & \textbf{Classification} & \textbf{Project life cycle} & \textbf{On-chain Patterns} & \textbf{AI Methods} & \textbf{Comp. w/ CeFi} & \textbf{Future Opportunities}\\
\hline
Badawi and Jourdan (2020) \cite{badawi2020cryptocurrencies} & General cryptocurrencies & By frauds &  $\times$ &   $\times$  &  \checkmark &  $\times$&  $\times$\\
\hline
Li et al. (2020) \cite{li2020survey} & General cryptocurrencies & By detection methods & N/A &  N/A & \checkmark & $\times$ &\checkmark \\
\hline

Eigelshoven et al. (2021) \cite{eigelshoven2021cryptocurrency} & General cryptocurrencies & By frauds &  $\times$ & \checkmark &  $\times$ & \checkmark & \checkmark \\
\hline
Bartoletti et al. (2021) \cite{bartoletti2021cryptocurrency} & General cryptocurrencies & By frauds &  $\times$& \checkmark &  $\times$ &  $\times$ & \checkmark   \\
\hline
Trozze et al. (2022) \cite{trozze2022cryptocurrencies} & General cryptocurrencies & By frauds & $\times$ &  $\times$ &  $\times$ & $\times$ & \checkmark   \\
\hline
Li (2022) \cite{li2022survey} & Ethereum & By detection methods &  N/A & N/A & \checkmark &  $\times$ & \checkmark  \\
\hline
Yan et al. (2022) \cite{yan2022blockchain} & Public and consortium blockchain & By frauds & $\times$ &  $\times$ & \checkmark &  $\times$ & \checkmark  \\
\hline

Wu et al. (2023) \cite{wu2023financial} & Web3-metaverse & By frauds  & $\times$ & $\times$   &  $\times$ & \checkmark & \checkmark \\
\hline
Krishnan et al. (2023) \cite{krishnan2023scams} & General cryptocurrencies & By frauds & $\times$ &  $\times$ & \checkmark &  $\times$ & $\times$ \\
\hline
Our Survey & DeFi & By frauds & \checkmark & \checkmark & \checkmark & \checkmark & \checkmark\\
\hline
\end{tabular}}
\label{table:review-comparison}
\end{table}

As shown in \autoref{table:review-comparison}, the existing literature has certain limitations: (1) Most studies have focused on general cryptocurrency markets, though a few have covered some frauds employed in Ethereum smart contracts. They have not, however, investigated other critical DeFi components. As discussed in \Cref{sec:defi_crypto}, it is essential to recognize that the emerging DeFi ecosystem encompasses a broader range of financial components, which have their own unique features and challenges in comparison with cryptocurrency. (2) There is no systematic taxonomy for different types of frauds. Most studies surveyed specific frauds, but there was no comprehensive categorization of them, which restricted a thorough understanding. (3) While detection methods are discussed in some surveys, there is a lack of in-depth analysis of on-chain patterns associated with these frauds and meaningful comparisons with CeFi. 

Our survey aims to address these limitations and deepen our understanding of the evolution of DeFi frauds. The main contributions of this paper are as follows: (1) We present a comprehensive analysis of DeFi frauds, including their on-chain patterns, characteristics, and existing typical and state-of-the-art detection methods, to aid researchers and practitioners. (2) We are the first to provide a taxonomy of DeFi frauds based on the life cycle of a DeFi project, resulting in better classification and understanding of fraud distribution in the DeFi ecosystem. (3) We compare CeFi and DeFi, including their characteristics, corresponding frauds and detection methods, to enhance our comprehension of the financial ecosystem. (4) We present intriguing research directions in DeFi fraud detection and regulation challenges that can guide researchers and regulators for future advancements.

\section{Overview}
\label{sec:overview}

\subsection{DeFi Fraud Taxonomy}
When surveying the frauds in DeFi, we discover that there are many different types of frauds, such as Ponzi schemes \cite{vasek2015there}, money laundering \cite{weber2019anti}, and pump and dump schemes \cite{xu2019anatomy}. Given various fraud types, previous surveys \cite{badawi2020cryptocurrencies, trozze2022cryptocurrencies} have focused on specific fraud types but have failed to provide a systemic review of them. 

When further analyzing the occurrence time of frauds within a DeFi project, we discover that the types of frauds have strong and consistent correlations with the DeFi project life cycle, regardless of bootup times, applications, or purposes. We follow the Product Life Cycle Theory \cite{SurveyMonkey} that outlines five crucial stages: development, introduction, growth, maturity, and decline. To illustrate this finding, we list some notable examples below: 

\begin{enumerate}
    \item Many Ponzi schemes and rug pulls emerging across a variety of projects, including coins, NFT, and exchanges, are designed to be fraudulent from the start. For example, the "Squid Game" coin enticed investors but vanished with millions\footnote{\url{https://www.wired.co.uk/article/squid-game-crypto-scam}. Accessed on Aug 16, 2023.}. Similarly, the Frosties NFT garnered funds through minting then disappeared abruptly\footnote{\url{https://nftnow.com/news/two-charged-in-frosties-pfp-nft-rug-pull/}. Accessed on Aug 16, 2023.}. 
    \item Another time-point prone to various frauds is around the launch of projects. For instance, the SEC charged insiders with trading crypto assets just before their official Coinbase availability\footnote{\url{https://www.steptoe.com/en/news-publications/blockchain-blog/secs-insider-trading-complaint-places-the-entire-defi-and-crypto-industry-in-a-bind.html}. Assessed on Aug 16, 2023.}. Wash trading also emerged during the inception of many NFT collections like Meebits, Rollbots, and OGCrystals \cite{la2022nft}.
    \item As DeFi projects mature, they encounter various other fraudulent activities. For instance, major cryptocurrencies such as Bitcoin and Ethereum become avenues for phishing \cite{chen2020phishing} and money laundering \cite{chainalysis2022moneylaundering}. Meanwhile, low market-cap coins like BVB, CON, and MAGN are targeted for pump and dump scams after maturing \cite{xu2019anatomy}. 
\end{enumerate}

\begin{figure}
\setlength{\abovecaptionskip}{0cm}
\setlength{\belowcaptionskip}{0cm}
\centering 
\includegraphics[scale=0.68]{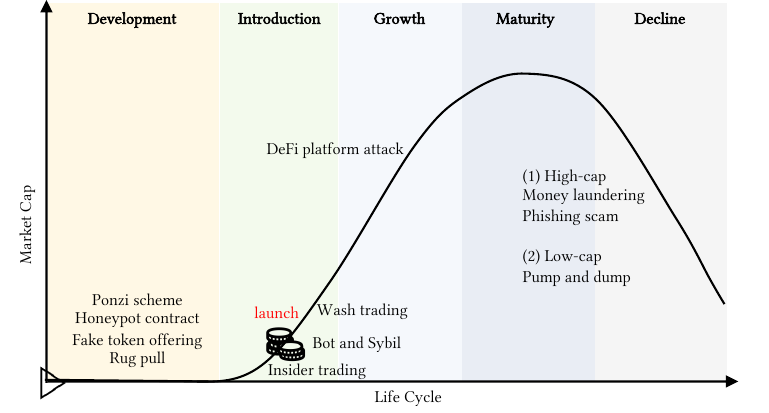} 
\caption{DeFi Frauds along Project Life Cycle.} 
\label{fig:life cycle} 
\vspace{-1em}
\end{figure}

Given the evident correlation between the occurrence of frauds and the various stages of DeFi projects, it is imperative to establish a comprehensive DeFi fraud taxonomy along the project's life cycle. Therefore, we present a taxonomy followed the stages of the DeFi project, as shown in \autoref{fig:life cycle}. A lot of common fraud types occur in introduction and growth stages, aligning with observations in maturity and decline stages. Hence, we group introduction and growth stages, as well as maturity and decline stages. We highlight the specific frauds associated with each stage as follows:

\begin{enumerate}
\item \textbf{Development:} During the development stage, fraudulent projects including Ponzi schemes, honeypot contracts, fake token offerings, and rug pulls are designed to deceive investors. These projects often collapse shortly after launching, resulting in significant losses for investors.

\item \textbf{Introduction \& Growth:} The introduction and growth stages represent project launch and subsequent rapid expansion. Concerns have arisen regarding insider trading, the activities of bot and Sybil accounts, and wash trading, all of which manipulate a project's market dynamics. As project value and user numbers increase, attackers exploit vulnerabilities in DeFi platforms like front-running and flash loan attacks.

\item \textbf{Maturity \& Decline:} As a project progresses into the maturity and decline stages, a different set of financial frauds come into play. High-cap projects become attractive targets for various financial crimes, including money laundering and phishing scams. Conversely, low-cap projects are vulnerable to pump and dump schemes due to their small market capitalization, making them relatively easy to manipulate.
\end{enumerate}

Moreover, it is important to note a few things of our analysis scope. First, while certain frauds can manifest at any project stage, we primarily focus on the stage when they most likely to occur. Second, our analysis centers on the primary financial frauds in DeFi, spanning smart contracts, DeFi platforms, DEXs, etc. We also delve into emerging cryptocurrency frauds that take on distinct forms compared to CeFi utilizing cryptocurrencies and other DeFi tools. Nevertheless, we exclude certain ones like darknet activity and ransomware, which only use cryptocurrencies, mainly Bitcoin, as payment tools without involving other DeFi components. 

Our taxonomy can be used to differentiate between various types of frauds based on their occurrence at various stages of DeFi projects. Distinct on-chain patterns and characteristics distinguish these various types of fraud during these stages, emphasizing the critical importance of deploying effective detection methods.

\subsection{AI-powered Techniques}
With the rapid growth of DeFi markets and diverse fraud types at different project stages, traditional manual fraud detection methods struggle to keep pace. As a result, researchers have leveraged more advanced AI-powered techniques to detect the DeFi frauds, such as statistical techniques, natural language processing (NLP) methods, and various other machine learning models, etc. In this subsection, we provide an overview of these techniques.

First, statistical techniques such as statistical modeling and statistical test are widely utilized to explore market dynamics and identify fraudulent activities \cite{cong2021crypto, suspicious-NFT, zwang2018detecting}. For instance, regression analysis \cite{freund2006regression} helps identify significant variables contributing to frauds. Multiple statistical tests, such as Benford's Law and Power Law Distribution test, detect unusual spikes in transaction volumes and activities in DeFi. These techniques are straightforward, intuitive, and easy to understand. Furthermore, unlike machine learning models, they do not require large amounts of data. They are especially common in addressing frauds that occur during the early stages of projects.

Second, NLP techniques play an significant role in analyzing the textual content of DeFi projects \cite{bian2018icorating, toma2020initial} and the smart contracts codes \cite{chen2021sadponzi, chen2021improving}. These techniques entail the analysis and processing of various textual information, such as DeFi project details, websites, and social media posts. They provide valuable insights that aid in project evaluation and sentiment analysis within textual content. Furthermore, the use of NLP techniques extend to smart contracts, where methods like TF-IDF and semantic-aware detection are employed to extract crucial information from contract codes. These techniques aid in the understanding of DeFi project documents as well as the effective identification of fraudulent contracts. They enable effective fraud detection prior to project launch and without the use of transactions.

Third, various machine learning models, including supervised and unsupervised learning, are valuable for various fraud detection tasks \cite{sun2022lstm, li2022ttagn, hu2023sequence, kumar2020edarkfind}. Supervised algorithms are trained on labeled datasets to identify suspicious accounts, fraudulent transactions, and predict fraud-targeted projects. Traditional models like Random Forest (RF) \cite{rigatti2017random} and XGBoost \cite{chen2015xgboost} achieve good performance and offer interpretability, while deep learning models like Long Short-Term Memory (LSTM) \cite{hochreiter1997long} and effectively analyze temporal data and sequential patterns of fraud. Graph-based techniques model DeFi transaction networks as large graphs, applying graph-embedding and well-designed Graph Neural Network (GNN) models \cite{zhou2020graph} to aid in detecting fraudulent activities. Moreover, unsupervised and self-supervised learning, such as clustering algorithms \cite{madhulatha2012overview}, uncover underlying patterns and anomalies in the data. Various machine learning models have been designed and adapted to detect frauds throughout DeFi project life cycle, with notable performance achieved using rich on-chain and off-chain data.

In summary, AI-powered methods have made significant contributions to DeFi fraud detection efforts. In the following sections, we examine various types of fraud and discuss the mentioned AI methods used to detect them at each stage. We also discuss the common characteristics of these techniques at each stage, given the strong correlation between fraud occurrence stages and detection methods.

\section{Fraud Detection in the Development Stage}
\label{sec:development}
During the development phase of DeFi projects, scams can be designed right from the start. These fraudulent schemes include Ponzi schemes, honeypot contracts, fake token offerings, and rug pulls. In this section, we will provide a thorough analysis of each type of fraudulent activities, including their patterns, related DeFi components, characteristics, and specific techniques for detecting and preventing them. An overview is presented in \autoref{tab:defi-fraud-detection1}. 

\begin{table}[htbp]
\centering
\setlength{\abovecaptionskip}{0cm}
\setlength{\belowcaptionskip}{-0.2cm}
\caption{Summary of DeFi Frauds and Detection Methods in the Development Stage. 
It includes on-chain patterns, targeted DeFi components or cryptocurrencies, and the relevant papers for each type of fraud.}
\resizebox{\textwidth}{!}{
\begin{tabular}{m{2.5cm}<{\centering} m{4.3cm}<{\centering} m{3.8cm}<{\centering} m{6.2cm}<{\centering}}
\hline
\textbf{Fraud Type} & \textbf{Patterns} & \textbf{Related Components} & \textbf{Papers} \\
\hline
\multirow{4}{*}{Ponzi scheme} & High-yield investment program (HYIP) & Bitcoin & \cite{vasek2019analyzing, Bartoletti_2018, vasek2015there, toyoda2017identification} \\
\cmidrule(lr){2-4} 
 & DeFi projects employed using smart contracts & Smart contracts & \cite{Bartoletti_2020, Jung_2019, Chen_2019, fan2022smart, Peng_2020, chen2021improving, chen2021sadponzi, zhang2021code, Zhang_2022, Zheng_2022, Fan_2021, Wang_2021, jin2022heterogeneous, chen2018detecting, Jin_2022, lu2023sourcep, wang2023detecting, wen2023code} \\
\hline
Honeypot contract & Traps in contracts & Smart contracts & \cite{torres2019art, camino2019data, chen2020honeypot} \\
\hline
Fake token offering & Counterfeit tokens & Tokens & \cite{liebau2019crypto, toma2020initial, bian2018icorating, karimov2021identification, meoli2022machine, zheng2022unravelling} \\
\hline
\multirow{3}{*}{Rug pull} & Scam tokens and liquidity pools & DEXs & \cite{xia2021trade, mazorra2022not, cernera2022token, nguyen2023rug} \\
\cmidrule(lr){2-4} 
& Scam domains and fake apps & Exchanges & \cite{xia2020characterizing} \\
\cmidrule(lr){2-4} 
 & Exit scams & NFTs, yield farms, etc. & \cite{agarwal2023short,sharma2023understanding, huang2023deep} \\
\hline
\end{tabular}
}
\label{tab:defi-fraud-detection1}
\end{table}

\subsection{Ponzi Scheme}
Ponzi schemes are fraudulent schemes in which new investors' funds are used to repay previous investors \cite{frankel2012ponzi, artzrouni2009mathematics}. In DeFi, Ponzi schemes were first known as "Bitcoin-only High Yield Investment Programs (HYIPs)" in 2013 \cite{MukherjeeLarkin}. With the advent of smart contracts making blockchain platforms more accessible, these schemes evolved with new patterns emerging within the contracts \cite{chen2018detecting}. These schemes collapse when new investments are insufficient to fulfill returns to earlier investors, with many lasting less than a week on Bitcoin and close to 0 days on Ethereum \cite{Bartoletti_2020, vasek2019analyzing}. 

\subsubsection{Patterns of Ponzi Contracts}

Understanding the patterns of Ponzi schemes is essential for detecting them. \autoref{fig:ponzi} shows the common patterns of Ponzi schemes deployed in the smart contracts \cite{Bartoletti_2020, chen2021sadponzi}, which include: 

\begin{figure}
\centering 
\setlength{\abovecaptionskip}{0cm}
\setlength{\belowcaptionskip}{0cm}
\includegraphics[scale=0.5]{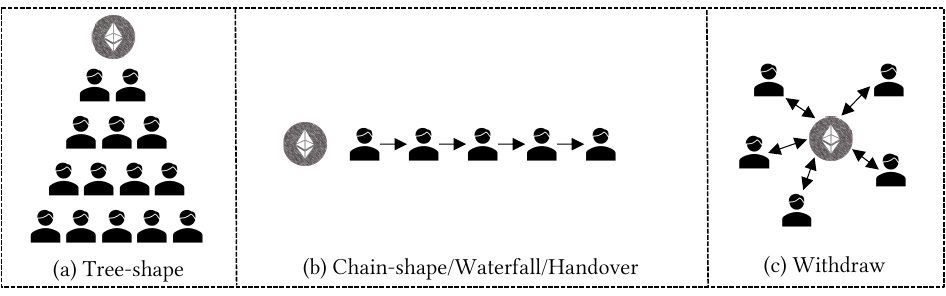} 
\caption{Common Patterns in Ponzi Contracts.} 
\label{fig:ponzi} 
\vspace{-1em}
\end{figure}

\begin{itemize}
\item \textit{Tree-shape}: New investors typically designate a parent node, and their investment is then divided among ancestors using a logic that assigns a larger share to closer ancestors.
\item \textit{Chain-shape}: Each parent node only has one child. The payout is fixed, and whenever a new child is added, it pays out to as many of the early investors as possible.
\item \textit{Waterfall}: Similar to a chain, but new investments are distributed along a chain of investors, with each participant receiving a share, and the distribution begins at the beginning of the chain.
\item \textit{Handover}: Similar to a chain, but with a predetermined entry toll that increases with new investors and is paid in full by previous investors.
\item \textit{Withdraw}: Similar to an Ether pool that grows through investments and is then distributed among participants. Investors in this scheme need to call the withdraw function to collect their funds.
\end{itemize}

Among these patterns, the chain-shape pattern has been identified as the most prevalent \cite{Bartoletti_2020}, possibly due to its straightforward and predictable payout structure, which attracts both investors and fraudsters. Additionally, there are other existing schemes, such as being experimental and variations of existing schemes.

\subsubsection{Detection Methods of Ponzi Schemes}
After developing a deep understanding of the patterns of Ponzi schemes, scholars have detected them manually from forums and developed various detection methodologies using AI methods.

(1) {\it Manual Analysis - }
Experts conduct in-depth investigations on forums and smart contract codes to identify suspicious activities. For instance, Vasek and Moore \cite{vasek2015there} manually analyzed Bitcoin Talk\footnote{\url{https://bitcointalk.org/}. Accessed on Aug 16, 2023.} and subreddit\footnote{\url{https://www.reddit.com/r/Bitcoin/}. Accessed on Aug 16, 2023.} threads, categorizing Bitcoin frauds, including Ponzi schemes. They expanded by studying more Bitcoin Talk threads to identify Ponzi schemes targeting Bitcoin users \cite{vasek2019analyzing}. Similarly, Bartoletti et al. \cite{Bartoletti_2020} manually inspected Ethereum smart contract codes, revealing hidden Ponzi schemes using techniques like normalized Levenshtein distance \cite{yujian2007normalized}.

(2) {\it AI Detection Methods - }
In order to overcome scalability and subjectivity limitations, researchers have turned to AI methods to detect Ponzi schemes. For Bitcoin Ponzi detection, Toyoda et al.\cite{toyoda2017identification} employed XGBoost and RF models to identify HYIP-associated Bitcoin addresses using transaction-based features, enhancing detection accuracy. Bartoletti et al. \cite{Bartoletti_2018} further improved detection in Bitcoin using cost-sensitive Random Forest classifiers based on \cite{toyoda2017identification}.

For Ethereum Ponzi contracts, researchers built machine learning models using rich features including account, opcode, developer information, and other information extracted by NLP techniques. Chen et al. \cite{chen2018detecting} started by building XGBoost models that utilizes account and opcode features, highlighting their importance for the model's performance. Later, they expanded their dataset and adopted RF as a bagging-based algorithm \cite{Chen_2019}. More related research has enriched feature sets by incorporating behavior-based features \cite{Jung_2019}, account interaction features \cite{fan2022smart}, developers information \cite{Zheng_2022}, and applying NLP techniques \cite{Peng_2020} such as TF-IDF and novel semantic-aware methods \cite{chen2021sadponzi, chen2021improving} for feature extraction. Additionally, researchers addressed imbalanced dataset and prediction shift issue. Techniques like oversampling-based LSTM models \cite{Wang_2021}, oversampling-based AdaBoost \cite{wang2023detecting}, anti-leakage detection \cite{Fan_2021} are applied, while transaction graph modeling in HFAug \cite{jin2022heterogeneous}, and Ponzi-Warning \cite{Jin_2022} are utilized as augmentation strategies. GNN models have also been utilized in \cite{jin2022heterogeneous, Jin_2022} to detect Ponzi schemes within transaction graphs. Notably, Lu et al. \cite{lu2023sourcep} recently designed SourceP, utilizing pre-trained models for code representation learning and solely relying on source code features for Ponzi contract detection, while Wen et al. \cite{wen2023code} created visual tools for identifying Ponzi contracts.

\begin{figure}
\centering 
\setlength{\abovecaptionskip}{0cm}
\setlength{\belowcaptionskip}{0cm}
\includegraphics[scale=0.55]{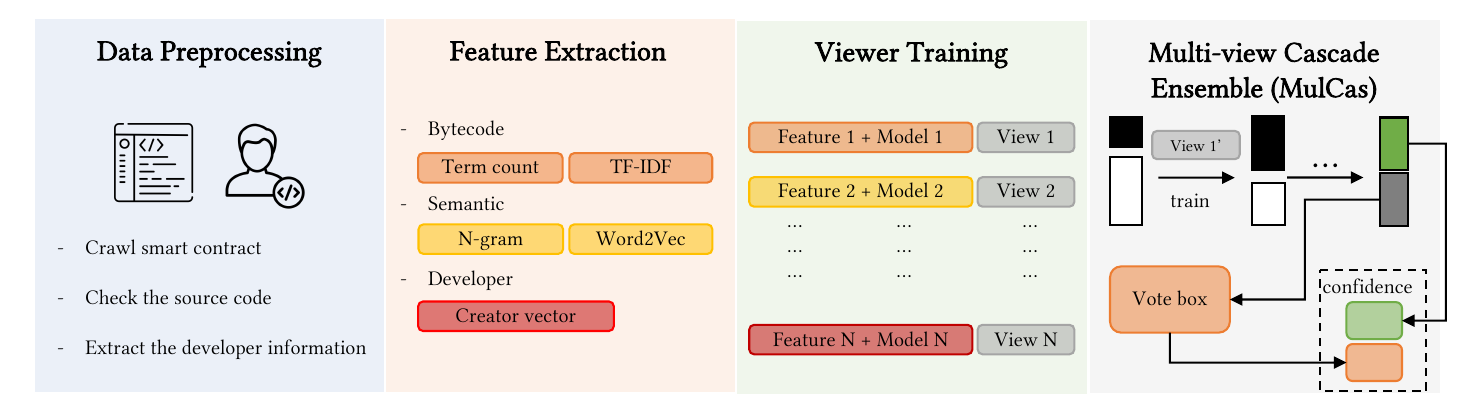} 
\caption{Process of Detecting Ponzi Scheme (Adapted from \cite{Zheng_2022}).} 
\label{fig:ponzi-scheme} 
\vspace{-1em}
\end{figure}

As an illustrative example, we further illustrate the state-of-the-art model proposed by Zheng et al. \cite{Zheng_2022}. \autoref{fig:ponzi-scheme} provides an overview of the model. The researchers collected 6,498 smart contracts, including 314 Ponzi contracts from Etherscan\footnote{\url{https://etherscan.io/}. Accessed on Aug 16, 2023.}. After that, they constructed multiple layers of rich features, including bytecode features (term count, TF-IDF) and semantic features (N-gram, Word2Vec), along with developer information (a creator vector flags previous Ponzi contract creators). These features enable the model to perform real-time classification during the creation of smart contracts. After building these features, the researchers utilized various classification models, including SVM, Ridge Regression, RF, and XGBoost, selecting the top-performing models for each type of features on the test set. With the selected models, they proposed the Multi-view Cascade Ensemble model (MulCas), incorporating a cascade ensemble and a vote-based ensemble for training and predicting, respectively. The model showed remarkable performance and demonstrated robustness on the imbalanced dataset.

\emph{Summary.} Scholars have employed rich data features for detection Ponzi schemes in DeFi. However, limitations persist when detecting newly launched or limited transaction history Ponzi schemes using features beyond contract codes and developer information. To address this, scholars can concentrate on advanced techniques that emphasize information available before the formal launch, enabling early detection.

\subsection{Honeypot Contract}
\subsubsection{Characteristics of Honeypot Contract}
Honeypot contracts have gained significant attention in Ethereum since 2018 \cite{torres2019art}. These contracts entice users with promises of high returns or special privileges. One notable example is the reported contract "0x8B3e6E910dfD6B406F9f15962b3656E799f60d2B"\footnote{\url{https://etherscan.io/address/0x8b3e6e910dfd6b406f9F15962b3656e799f60d2b}. Accessed on Aug 16, 2023.} \cite{Oualid_2022}. In the source code, the "multiplicate" function promises to provide the investor's contributed value and the current balance. However, a careless user may overlook the fact that the contract's balance increases when the function is executed, resulting msg.value to be consistently lower than the balance. As a result, the investor will never receive the transfer.

\begin{verbatim}
function multiplicate(address addr) public payable {
    if(msg.value >= this.balance) {
        addr.transfer(this.balance + msg.value);
    }}
\end{verbatim}

Research has shown that honeypot contracts, similar to Ponzi schemes, have a remarkably short average life cycle \cite{torres2019art}. On average, the median life cycle of a honeypot is around 3 days following their exploitation. This short life cycle emphasizes the importance of developing an early-detection system to prevent this scam.

\subsubsection{Detection Methods of Honeypot Contracts}
Limited research has been focused on the detection of honeypot schemes. Torres and Steichen \cite{torres2019art} conducted a systematic analysis of honeypot contracts and introduced HoneyBadger, a tool that utilizes symbolic execution and heuristics for automated honeypot detection. Building upon this research, Camino et al. \cite{camino2019data} developed an AI method that incorporated source code information, transaction properties, and fund flows as features and used XGBoost as classification model. Further advancements were achieved by Chen et al. \cite{chen2020honeypot}, who obtained 857 honeypot contracts from the HoneyBadger \cite{torres2019art} project and extracted N-gram features from the contracts' opcode. They utilized LightGBM \cite{ke2017lightgbm}, to classify the honeypot contracts. Given the significant class imbalance of the dataset, the authors applied undersampling of negative samples multiple times. They constructed a model in each iteration using sampled negative and positive samples, which ultimately achieved impressive results.

\emph{Summary.} Despite these notable efforts, the detection of honeypot contracts remains limited, possibly due to insufficient contract codes or the overlap with detecting honeypot contracts. Thus, scholars have the opportunity to investigate the potential transferability of models designed to detect Ponzi schemes to enhance this field.

\subsection{Fake Token Offering} 
\subsubsection{Definition of Fake Token Offering} 
Fake token offerings refer to the creation and promotion of counterfeit tokens that deceive investors by failing to fulfill project development obligations with the raised funds \cite{liebau2019crypto}. In related studies, scholars have identified fake offerings using different approaches. One approach focused on detecting the absence of social media activities from the tokens' official accounts and their delisting from listing platforms, which serve as red flags indicating that the tokens no longer exist \cite{karimov2021identification}. Other approaches involved analyzing token performance, such as significant price declines shortly after a successful initial offering or abnormal fundraising performance \cite{bian2018icorating, meoli2022machine}. 

\subsubsection{Detection of Fake ICO} 
The identification of fake ICO has been a focus in the field. Statistical analysis has been widely employed to uncover the characteristics and impact of scams in ICOs. Liebau and Schueffel \cite{liebau2019crypto} conducted an empirical study using Principal Agent Theory and statistical analysis to establish the current state of scam ICOs. Toma and Cerchiello \cite{toma2020initial} employed statistical methods like Logistic Regression (LR), Multinomial Logistic Regression, and text analysis to study factors influencing ICO success. They found key indicators of fraudulent behavior, including the presence of a website, whitepaper, Twitter account, and sentiment in Telegram chats. 

Machine learning techniques have also been applied to detect scam ICO projects. Bian et al. \cite{bian2018icorating} proposed IcoRating, a machine learning-based cryptocurrency rating system, to identify scam ICO projects. They extracted features from white papers, funding teams, websites, and GitHub repositories of 2,251 ICO projects and applied neural network models to classify scam ICOs. As shown in \autoref{case_y}, their models used the price changes of ICO projects over one year as training signals. They categorized a project suspicious if the predicted price falls below m percent of its ICO value. Evaluation shows that when setting m = 1, the model achieved best performance.
\begin{equation} \label{case_y}
  y=\begin{cases} 
       1 & \text{if } \frac{\text{price}(t)}{\text{price}(0)} \leq m \text{, t = a year} \\
       0 & \text{otherwise.}
    \end{cases}
\end{equation}

{\noindent}Additionally, several studies have employed various machine learning models, including SVM, Naive Bayes, and tree-based models, etc. to detect unsuccessful ICO projects \cite{karimov2021identification, meoli2022machine}. Another related work examined the EOSIO token ecosystem, identifying fake tokens based on users' account-creation relationships \cite{zheng2022unravelling}.

\emph{Summary.} While statistical analysis and machine learning models have proven useful in detecting fake ICO scams and unsuccessful ICOs, there is a need for further research that explores other types of fake token offerings, such as fake IDOs and fake airdrops. Moreover, researchers should standardize the fake ICO definition to compare dataset reliability and model performance. They can evaluate various criteria on real cases to determine the most appropriate one.

\subsection{Rug pull} 
Rug pulls, also known as exit scams, have gained significant attention since 2021 \cite{Chainalysis2022}. These scams involve the sudden disappearance or abandonment of DeFi projects, leaving investors with worthless assets. Scholars have studied the rug pulls target various components of DeFi, such as scam tokens in DEXs, NFTs, exchanges, and others.  

\subsubsection{Rug Pulls on Scam Tokens}
Scam tokens and their liquidity pools are a focus of this field. Xia et al. \cite{xia2021trade} identified scam tokens on Uniswap based on their name similarity to popular tokens. They proposed methods for detecting scam tokens using guilt-by-association and machine learning models, with RF outperforming the others. Mazorra et al. \cite{mazorra2022not} extended this work by categorizing rug pulls into types and using XGBoost and FT-Transformer with attention to classify the scam tokens. Further, Cernera et al. \cite{cernera2022token} extended the study to BSC, identifying rug pulls, token spammers, and sniper bots on both Ethereum and BSC DEXs. Recently, Nguyen et al. \cite{nguyen2023rug} proposed a new heuristic based on the factors for token price depreciation and classified 26,083 tokens as scam tokens and 631 normal. They constructed various features related to token, event, pool, creator, and average block-dependent features, and used XGBoost to detect rug pulls in Uniswap and learnt the most important features using ANOVA-based techniques.

\subsubsection{Rug Pulls on Other DeFi Components}
Rug pulls can also manifest as exchange scams, in which deceive users with promises of higher returns or bonuses \cite{metamask}. Xia et al. \cite{xia2020characterizing} conducted a study on cryptocurrency exchange scams and identified rug pull cases involving scam domains and fake apps. They revealed the relationships between these scam domains and fake apps, identifying multiple families of them. 

Additionally, scholars have also investigated rug pulls as exit scams across other categories of DeFi projects. Agarwal et al. \cite{agarwal2023short} collected and analyzed 101 rug pull cases on various DeFi projects, such as yield farms, NFTs, and cloud mining, from discussion forums. Sharma et al. \cite{sharma2023understanding} identified rug pull schemes across ten NFT marketplaces, exploring the structural and behavioral properties of these scams. Recently, Huang et al. \cite{huang2023deep} conducted a comprehensive study on NFT rug pulls. They first summarized NFT rug pulls tricks on 253 known cases and employed a rule-based method to identify 7,487 additional cases. Then, they developed a predictive model using LR, SVM, and RF with time-series data, token transfer logs, and secondary market trades features to proactively detect NFT rug pull projects. LR and SVM yielding impressive results in 4 days before the rug pulls.

\emph{Summary.} Scholars have conducted some research on rug pull scams in the DeFi ecosystem, with a particular focus on scam tokens and liquidity pools. However, there remains a need to develop more effective AI models for early-stage detection, specifically for rug pulls tailored on exchanges, NFTs, yield farms and others.

\subsection{Summary and Discussion of Development Stage Fraud Detection}
Through the review of fraud detection in the project development stage, we have the following findings.

\begin{itemize}
    \item NLP techniques are widely used in this stage. (1) To identify smart contract fraud, researchers leverage in-depth analysis on the contract codes. Although a few studies have introduced semantic methods, we expect more specific code review and analysis tools tailored to smart contracts. (2) NLP techniques are also utilized to analyze project websites, social platforms, and whitepapers to identify fake projects that may not have been officially launched or have limited transactions. However, confirming the alignment between project whitepapers and smart contract implementation remains complex, potentially leading to misleading investors commitments.

    \item Tree-based models have demonstrated superior performance compared to other models in this stage. With their robustness and interpretability, these models enable researchers to achieve good performance across diverse scenarios and explore feature significance. Moreover, given the limited early-stage DeFi ecosystem and associated frauds, tree-based models hold an advantage by requiring less training data than deep learning approaches, making them more appropriate for use when data is scarce \cite{grinsztajn2022tree}. 

    \item While a few studies implemented augmentation and over-sampling strategies, there is a lack of well-developed self-supervised or unsupervised methods for dealing with the scarcity of data for fraud detection in the project development stage. We expect further research in this promising field.
    
\end{itemize}

In general, the detection of frauds in the developmental phases of DeFi projects remains relatively nascent, particularly for fraud types beyond Ponzi schemes. They require more attention from researchers and practitioners.

\section{Fraud Detection in the Introduction \& Growth Stage}
\label{sec:introduction_growth}
In this section, we examine the fraudulent activities that commonly occur during the introduction and growth stages of a project when it launches and starts to grow rapidly. These activities include insider trading, the activities of Bot and Sybil accounts, wash trading and some attacks on DeFi platforms. An overview is provided in \autoref{tab:defi-fraud-detection2}.

\begin{table}[htbp]
\centering
\setlength{\abovecaptionskip}{0cm}
\setlength{\belowcaptionskip}{-0.2cm}
\caption{Summary of DeFi Frauds and Detection Methods in the Introduction \& Growth Stage.}
\resizebox{\textwidth}{!}{%
\begin{tabular}{m{2.5cm}<{\centering} m{7.2cm}<{\centering} m{3.2cm}<{\centering} m{3.5cm}<{\centering}}
\hline
\textbf{Fraud Type} & \textbf{Patterns} & \textbf{Related Components} & \textbf{Papers} \\
\hline
Insider trading & Consistent trading patterns prior to announcements & Exchanges & \cite{felez2022insider} \\
\hline
\multirow{2}{*}{Bot \& Sybil account} & Bot: Non-human entities & Across markets & \cite{zwang2018detecting, li2023understanding}\\
\cmidrule(lr){2-4} 
& Sybil: an entity creating numerous fake identities & Across markets & \cite{kumar2020edarkfind}\\
\hline
Wash trading & Various patterns involving buyer and seller from the same parties & Exchanges, tokens, NFTs & \cite{cong2021crypto, suspicious-NFT, bonifazi2023performing, le2021wash, chen2022cryptocurrency, tan2023bubble, aloosh2019direct, victor2021detecting, cui2022wteye, von2022nft, das2021understanding, serneels2022detecting, wen2023nftdisk, la2022nft, gan2022understanding}\\
\hline
\multirow{6}{*}{DeFi platform attack} & Front-running: deceptive tactics in transaction orderings & DEXs, DApps, coins & \cite{eskandari2020sok, daian2020flash, torres2021frontrunner, zhou2021high}\\
\cmidrule(lr){2-4} 
&  Wallet exploitation: steal assets from users' wallets & DeFi platforms and wallets &Notable examples are discussed.\\
\cmidrule(lr){2-4}
& Smart contract vulnerabilities: attacks due to the bugs in smart contracts &  DeFi platforms & Notable examples are discussed. \\
\hline
\end{tabular}%
}
\label{tab:defi-fraud-detection2}
\end{table}

\subsection{Insider Trading}
Insider trading \cite{wang2010insider}, observed in both CeFi and DeFi markets, involves individuals leveraging undisclosed information for trading advantage. In particular, in DeFi, it is likely to occur when certain individuals have non-public project information about a project. Many launchpads, found in both DeFi and CEX, participate in insider trading. For example, insider trading occurs notably before projects are listed on exchanges \cite{felez2022insider}. Law experts have delved into legal and regulatory nuances linked to this concern in DeFi \cite{verstein2019crypto, ur2019trust, travis2019common}.

However, the study of insider trading in the DeFi ecosystem has been relatively limited. Felez-Vinas et al. \cite{felez2022insider} conducted a notable study in which they uncovered evidence of systematic insider trading in cryptocurrency markets. By analyzing exchange announcements and transaction data across CEXs and DEXs from September 2018 to May 2022, the author detected abnormal return run-ups preceding official listing announcements, which strongly suggest insider trading. Additionally, they tracked wallets engaged in these tradings and estimated that insider trading occurs in approximately 10-25\% of cryptocurrency listings based on excess price run-ups, leading to trading profits of around \$1.5 million. However, we expect more studies focused on the on-chain patterns and detection of insider trading in the DeFi ecosystem. 

\subsection{Bot Accounts and Sybil Accounts}
During the introduction of DeFi projects, promotional activities can lead to a rise in unusual account actions. Notably, the existence of bot accounts is a significant concern. In DeFi, there are various types of bot accounts, such as arbitrage, market-making, and automated trading bots. To detect such bot activities in DeFi, Zhang and Somin \cite{zwang2018detecting} employ a Network Theory approach in the Ethereum blockchain network. They detected anomalies by examining whether the networks' degree distribution followed a power law pattern. This approach helped them identify non-human entities displaying periodic and irregular anomalies, particularly during significant events such as token airdrops. Another related work \cite{li2023understanding} analyzed the arbitrage bot contracts by collecting YouTube scam video creator accounts. However, there is a lack of comprehensive studies specifically focused on the direct detection of bot accounts or the classification and understanding of their behaviors.

Another form of suspicious account is the Sybil account, which involves an entity creating numerous fake identities for malicious purposes in the DeFi markets. For example, during the recent airdrop for the Ethereum scaling protocol Arbitrum\footnote{\url{https://www.coindesk.com/consensus-magazine/2023/04/10/crypto-airdrop-sybil-attacks/}. Accessed Aug 16, 2023.}, it was reported that nearly 48\% of all distributed tokens went to entities with Sybil accounts. Additionally, Sybil accounts can also refer to accounts belonging to the same entity across multiple blockchains or marketplaces. Kumar et al. \cite{kumar2020edarkfind} presented an unsupervised framework for detecting Sybil accounts among vendors across multiple darknet markets. Their model utilized various metaviews, including pre-trained language representation networks, domain-specific contextual information, stylometric features, location information, and substance information, achieving strong performance across different Darknet markets dataset. Given the limited research in this area, a research opportunity lies in utilizing advanced AI methods to identify Sybil accounts within individual chain or across different chains.

\subsection{Wash Trading}
Wash trading, observed in traditional stock markets, entails transactions between the same buyer and seller \cite{imisiker2018wash}. In DeFi, investigations since 2019 reveal similar patterns in token markets and cryptocurrency transactions \cite{aloosh2019direct, victor2021detecting, cui2022wteye}. Wash trading involving flash loans \cite{gan2022understanding} was also discovered. Furthermore, as the NFT market grows in popularity, wash trading becomes very likely to occur during the introduction phases of NFT collections or to earn trading rewards during the launching phases of new NFT marketplaces \cite{la2022nft, forkast}. Additionally, similar fraudulent activities are observed during the minting phase of NFTs, where creators of the collections may mint NFTs to deceive investors into believing the project is trending and investing in it\footnote{\url{https://nftgo.io/discover/hot-mint}. Accessed on Aug 22, 2023.}.

\subsubsection{Patterns of NFT Wash Trading}
\begin{figure}
\centering 
\includegraphics[width=0.77\linewidth]{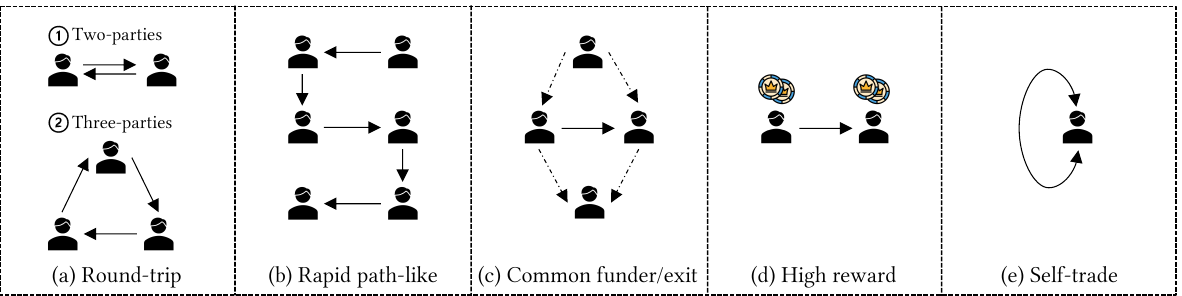} 
\caption{Common NFT Wash Trading Patterns (---> represents an NFT transaction, - -> represents an ERC20 transaction).} 
\label{fig:wash_trades} 
\vspace{-1em}
\end{figure}

Due to the unique identities of each NFT item, wash trading has exhibited more complex patterns in the NFT ecosystem. Scholars have identified various transaction patterns associated with potential wash trades as follows \cite{von2022nft, das2021understanding, la2022nft, duneanalytics}, illustrated in \autoref{fig:wash_trades}:

\begin{itemize}
    \item Round-trip transactions with matching-value or on a single NFT item. Two or more parties can take part.
    \item Path-like transactions consisting of rapid trade sequences.
    \item Transactions where buyers and sellers share the same funders (accounts that transferred funds to them before the transactions) or exits (funds transferred to the same account after the transactions).
    \item Transactions with exceptionally high rewards gained from marketplaces.
    \item Transactions with same buyer and seller. However, many exchanges have prohibited such self-trades \cite{coinbase_dex}.
\end{itemize}

While round-trip transactions remain the most obvious patterns in the NFT market, other notable ones have emerged, especially for buyers and sellers sharing the same funders or exits \cite{la2022nft}. The unique characteristics of the DeFi ecosystem contribute to the possibility of new wash trading patterns, making the detection of such activities more complex.

\subsubsection{Detection of Wash Trading}
Wash trading is difficult to detect in DeFi due to a lack of clear signals and reliable ground truth. To address this, statistical and graph techniques with expert-derived rules have been common used.

(1) {\it Statistic analysis -} Scholars employed statistical analysis to identify and examine the impact of wash trades in exchanges and other DeFi markets. For example, Cong et al. \cite{cong2021crypto} utilized statistical and trading patterns to detect wash trade transactions across multiple cryptocurrency exchanges. They conducted pooled regression analysis to estimate the proportion between trading volume that is not rounded and trading volume that is rounded, as shown in \autoref{eq:wash_trade}, where $V_{Unrounded_{it}}$ and $V_{Rounded_{it}}$ represent the unrounded and rounded volumes of regulated exchange $i$ in week $t$, respectively:
    \begin{equation}
    \label{eq:wash_trade}
    ln(V_{Unrounded_{it}}) = \alpha + \beta * ln(V_{Rounded_{it}}) + \gamma * X_{it} + \epsilon_{it}
    \end{equation}
According to their estimations, wash trading accounted for an average of over 70\% of the reported volume on each unregulated exchange. Similarly, Tariq and Sifat \cite{suspicious-NFT} employed statistical techniques such as Benford’s Law and clustering tests to identify suspicious wash trading activities in the NFT market. Later, Bonifazi et al. \cite{bonifazi2023performing} investigated NFT wash trades' profitability and found that engaging in such activities within the NFT market is not lucrative.

Moreover, researchers have devised statistical indicators to estimate the extent of wash trading. Le Pennec et al. \cite{le2021wash} distinguished accurately-reporting exchanges from those involved in wash trading based on web traffic metrics and user fund administration. Chen et al. \cite{chen2022cryptocurrency} defined a broader range of indicators by combining both on-chain and off-chain data to approximate wash trading volume. Tan et al. \cite{tan2023bubble} introduced statistical indicators like turnover ratio, price gaps, and concentration degree to identify wash trades in the NFT market.

(2) {\it Direct detection of wash trades -}
After conducting statistical analyses, researchers employed rules based on expert knowledge to directly detect wash trades. Aloosh et al. began by \cite{aloosh2019direct} identifying wash trading in Bitcoin transactions, referring to transactions where buyer and seller share the same trader ID. Then, to detect wash tradings in crypto tokens, Victor et al. \cite{victor2021detecting} and Cui et al. \cite{cui2022wteye} implemented graph-based algorithms, such as the detection of strongly connected components (SCCs) \cite{dorogovtsev2001giant} which represent strong relationships among traders. Gan et al. \cite{gan2022understanding} proposed a heuristic-based detection method on Ethereum transaction graphs for flash-loan based wash trading. Later, Victor et al. \cite{von2022nft} extended their detection to the NFT market by identifying topologically closed cycles or rapid trade sequences resembling paths. Others \cite{das2021understanding, serneels2022detecting, la2022nft} applied rules on SCC and cycle for more NFT wash trading detection. Recently, Wen et al. \cite{wen2023nftdisk} introduced NFTDisk, a visualization tool for spotting NFT market wash trading.

As an illustrative example, we delve into the state-of-the-art methodology proposed by La Morgia et al. \cite{la2022nft} for detecting wash trading activities in ERC721 NFTs. The authors collected all ERC-721 NFTs transactions from Ethereum's inception to Jan 2022, and constructed Directed MultiGraphs for each NFT item. These graphs represented transactions between accounts where each edge carried essential transaction information, such as timestamps, hashes, interacted smart contracts, and the values. After creating the graphs, the authors identified all SCCs in them and used four rules to detect wash trading: zero-risk position, common funder, common exit, and self-trade. They also confirmed more wash trading by identifying SCCs composed of the same nodes as previously detected SCCs. Their analysis revealed 12,413 confirmed wash trading activities in the NFT markets with a total volume of \$3 billion. After detection, they conducted comprehensive analyses on these activities, exploring their temporal patterns, serial wash traders, and profitability.

\emph{Summary.} These statistical analysis and graph-based techniques have identified some suspicious wash trading activities. However, when it comes to verifying the accuracy of detection algorithms or building robust machine learning models, the lack of reliable ground truth data poses significant challenges. Furthermore, DeFi users can easily create a secondary or multiple accounts, making the wash trading patterns more complex and difficult to detect.

\subsection{DeFi Platform Attack}
Many attacks on DeFi platforms, such as lending platforms and DEXs, emerge as the platform grows in popularity and more users begin to participate. Attackers may target the platforms' transaction orderings, wallets, and smart contracts.

\subsubsection{Front-running}
Front-running  \cite{eskandari2020sok} stands out among the vulnerabilities that impact DeFi platforms. In traditional market, front-running are similar to insider trading where individuals exploit early access to information \cite{bernhardt2008front}. However, in DeFi, it involved deceptive transaction ordering tactics, with attackers typically duplicating profitable transactions and using higher gas prices to displace, insert, or suppress victims' \cite{eskandari2020sok, torres2021frontrunner}. Notably, front-running vulnerabilities have been identified in DEXs \cite{daian2020flash}, DApps, and ICOs \cite{eskandari2020sok}. 

Heuristics and machine learning have been used to detect front-running. Torres et al. \cite{torres2021frontrunner} measured front-running attacks on Ethereum, detecting displacement, insertion, and suppression attacks by heuristics. Their analysis of over 11 million blocks revealed nearly 200K attacks, yielding attackers a total profit of \$18.41 million. Varun et al. \cite{zhou2021high} built on this work by employing transaction-related features and a multi-layer perceptron (MLP) model for front-running classification, achieving accuracy > 0.85 for all types of attacks. Given the scarcity of studies, advanced AI methods, particularly real-time detection methods with regards to the speed of front-running, are expected to develop.

\subsubsection{Other DeFi Platform Attacks}

Except for front-running, there are other notable vulnerabilities exist on the DeFi platform as well. We focus on two more important concerns, namely wallet exploitation and smart contract vulnerability. Although scholars have investigated the security aspects of cryptocurrency wallets \cite{Houy_Schmid_Bartel_2023} and smart contracts \cite{atzei2017survey, rouhani2019security}, the scarcity of these hacks and the diversity of exploitation methods have resulted in insufficient historical data for constructing statistical models or employing machine learning methods. Thus, we present some notable examples below instead of detection methods.

First, attackers may steal assets from wallets by attacking the platforms or software wallets. Notable examples include the Slope wallet hack\footnote{\url{https://www.coindesk.com/business/2022/08/03/solanas-latest-6m-exploit-likely-tied-to-slope-wallet-devs-say/}. Accessed on Aug 22, 2023.}, where \$ 6 million have been stolen from more than 9,000 hot wallets on Solana, and the recent Atomic wallet hack\footnote{\url{https://cointelegraph.com/news/north-korean-hackers-swipe-over-100m-from-atomic-wallet-users}. Accessed on Aug 22, 2023.}, where over \$100 million have been swiped over from about 5,500 crypto wallets. 

Second, attacks exploit vulnerabilities in smart contracts, whether existing or introduced through recent upgrades. For instance, Osmosis encountered a \$5 million loss due to a contract flaw enabling withdrawals 50\% higher than deposits\footnote{\url{https://finance.yahoo.com/news/osmosis-exploited-5m-due-contract-021049291.html}. Accessed on Aug 22, 2023.}. The Euler Finance flash loan attack\footnote{\url{https://cointelegraph.com/news/euler-finance-hacked-for-over-195m-in-a-flash-loan-attack}. Accessed on Aug 22, 2023.} also resulted from a smart contract bug, where the absence of health factor checks allowed the attacker to liquidate, repay the flash loan, causing a loss exceeding \$195 million.

Despite the limitation, given the huge losses they incur, practitioners and researchers need to devise safer platform designs and applicable countermeasures to prevent the recurrence of such attacks. For instance, developers can enhance safety by implementing thorough testing of smart contracts prior to deployment. Early-warning systems are also expected to protect platforms and users. Additionally, many common attacks are now thwarted by MEV (maximal extractable value). As a result, we also expect deeper exploration of MEV bot behaviors in this field.

\subsection{Summary and Discussion of Introduction \& Growth Stage Fraud Detection}
Through the review of fraud detection in the project introduction and growth stage, we have the following findings. 

\begin{itemize}
    \item Statistical techniques, including regression, statistical tests, and statistical indicators, are widely used in this phase due to limited data availability. These methods reveal correlations, identify abnormal market occurrences, and spot suspicious behaviors in DeFi. Further, more research on causal inferring across market sectors and fraudulent activities is also expected. 
    \item Graph-related fraud detection methods are also widely employed in this phase, which allow scholars to extract relationships between various accounts. Scholars detect special patterns based on rules, or examine whether the graph follows common distributions. In contrast, graph learning methods are rarely used during this phase.
\end{itemize}

In general, AI fraud detection methods have seen limited application in this phase, possibly due to a lack of transaction data in the early stages of projects or the difficulty of effective detection in the volatile DeFi markets. This opens up some possibilities for future work.

\section{Fraud Detection in the Maturity \& Decline Stage}
\label{sec:maturity_decline}
After launching, DeFi projects gradually mature and may start to decline. Different frauds can occur depending on the market cap of the projects in this phase. While high-cap projects may be vulnerable to money laundering and phishing scams, low-cap projects may be targeted for pump and dump schemes. An overview is provided in \autoref{tab:defi-fraud-detection}.

\begin{table}[htbp]
\centering
\setlength{\abovecaptionskip}{0cm}
\setlength{\belowcaptionskip}{-0.2cm}
\caption{Summary of DeFi Frauds and Detection Methods in the Maturity \& Decline Stage.}
\resizebox{\textwidth}{!}{%
\begin{tabular}{m{2.5cm}<{\centering} m{5.2cm}<{\centering} m{3.2cm}<{\centering} m{5.5cm}<{\centering}}
\hline
\textbf{Fraud Type} & \textbf{Patterns} & \textbf{Related Components} & \textbf{Papers} \\
\hline
Money laundering & Sophisticated techniques with wallets and mixers to obscure transactions & Cryptocurrencies, NFTs  & \cite{moser2013inquiry, shojaeenasab2022mixing, wu2021detecting, weber2019anti, lo2023inspection, lorenz2020machine, alarab2020competence, hu2019characterizing, oad2021blockchain, sun2022lstm, vassallo2021application} \\

\hline
\multirow{3.5}{*}{Phishing scam} & Websites (including giveaway scams) & Cryptocurrencies, NFTs & \cite{vakilinia2022cryptocurrency, li2023double, li2023understanding, saha2023demystifying, phillips2020tracing} \\
\cmidrule(lr){2-4} 
& Directly transfer to phishing accounts & Mainly Ethereum &\cite{yuan2020detecting, wu2020phishers, chen2020phishing, narayanan2017graph2vec, xia2022phishing, yuan2020phishing, wang2021tsgn, zhang2019graph, zhou2022behavior, li2022ttagn, shen2021identity, zhang2021blockchain, yu2022mp, fu2022ct, kanezashi2022ethereum, hu2023bert4eth, yu2023streaming, zhou2023detecting, wan2023early, ghosh2023investigating, wen2022hide, wen2021transaction} \\
\hline
Pump and dump & Rapid price change, closely related to social media activities & Mainly coins & \cite{xu2019anatomy, nghiem2021detecting, hu2023sequence, kamps2018moon, la2023doge, la2020pump, victor2019cryptocurrency, chen2019detecting} \\
\hline
\end{tabular}%
}
\label{tab:defi-fraud-detection}
\end{table}

\subsection{Money Laundering}
\subsubsection{Patterns of Money Laundering}
Money laundering is the act of presenting illegally obtained funds in a manner that gives the impression of their legitimacy \cite{schneider2008money}. Since the rise of Bitcoin's popularity in 2013, money laundering in the DeFi ecosystem has been widely studied. Economists \cite{brenig2015economic, dyntu2018cryptocurrency} have investigated the incentives that cryptocurrencies may provide for money laundering and the global anti-money laundering measures. Nowadays, privacy coins like Monero are also widely used for laundering, due to the anonymity they provide to users\footnote{\url{https://www.ft.com/content/13fb66ed-b4e2-4f5f-926a-7d34dc40d8b6}. Accessed on Aug 28, 2023}. 

In comparison to traditional financial markets, money laundering in DeFi has become more sophisticated. Criminals use a variety of entities and services to temporarily hold funds, conceal their movements, and facilitate cryptocurrency swaps, such as personal wallets, mixers, and darknet markets. Furthermore, they rely on fiat off-ramps, such as centralized exchanges, peer-to-peer exchanges, and other services, to convert cryptocurrencies into fiat currencies \cite{Team_2023}. Money laundering has also spread to NFT markets, where intermediaries may be used to purchase NFT items with illicit funds and then sell them to launder money \cite{jordanoska2021exciting}. 

\subsubsection{Detection of Money Laundering}
Researchers have employed various AI methods to study money laundering services and activities in DeFi.

(1) {\it Detecting laundering services and belonged accounts -} Understanding the use of money laundering in cryptocurrency markets necessitates a thorough examination of laundering services and the accounts associated with them. Möser et al. \cite{moser2013inquiry} explored anti-money laundering opportunities and limitations in Bitcoin, focusing on on-chain laundering services and analyzing the transaction graph for insights. Shojaeenasab et al. \cite{shojaeenasab2022mixing} employed statistical analysis to identify patterns in mixing transactions and addresses within the Bitcoin blockchain, investigating causes and illustrating the transaction chain involved in money laundering. In addition, Wu et al. \cite{wu2021detecting} employed machine learning models to detect suspicious addresses associated with mixing services. They collected ground truth data by extracting labels of three services, namely Bitcoin Fog, BitLaunder, and Helix, from WalletExplorer\footnote{\url{https://www.walletexplorer.com/}. Accessed on Aug 16, 2023.}. Subsequently, they identified network, account, and transaction-level statistical properties of mixing services and developed a positive and unlabeled (PU) learning model to enhance the model's performance. More recently, 

(2) {\it Detecting money laundering transactions - } Moreover, various machine learning methods have been employed to detect and identify money laundering activities in DeFi. In 2019, Weber et al. \cite{weber2019anti} introduced the most widely used dataset in this field, the Elliptic Dataset. It correlates Bitcoin transactions with real-world entities and contains over 203k transactions with 234k directed edges. Among these transactions, 4,545 are labeled as illicit, 42,019 as licit, and the rest are categorized as undefined. Each transaction is associated with 166 features, including 94 local features such as transaction fee, timestep, and number of inputs/outputs, and 72 aggregated features derived from one-hop backward/forward from the central node, such as the standard deviation of neighboring transactions. With this dataset, the authors employed multiple traditional machine learning algorithms, such as LR, MLP, and RF, along with graph-related models like GCN and EvolveGCN, a temporal model that extends GCN, to identify illicit transactions. RF using all features and node embeddings computed by GCN achieved the best performance, while EvolveGCN also demonstrated good results. 

The Elliptic Dataset is widely employed in subsequent studies, with enhanced model performance achieved through novel-designed layers in GCN models \cite{alarab2020competence}, eXtreme Gradient Boosting model adaptations \cite{vassallo2021application}, and frameworks incorporating self-supervised GNN \cite{lo2023inspection}. Moroever, Lorenz et al. \cite{lorenz2020machine} introduced active learning solutions for Bitcoin money laundering detection with minimal labeled data access. Additionally, Hu et al. \cite{hu2019characterizing} compiled their own dataset by scraping abnormal transactions from major laundering services, including AlphaBay, BTC-e, Bitmixer, and HelixMixer. They employed graph techniques to detect money laundering in the Bitcoin transaction network using immediate neighbors, curated features, deepwalk embeddings, and node2vec embeddings features.

Recently, XBlockFlow, the first dataset of money laundering on the Ethereum network is presented \cite{wu2023towards} which extracted through the lenses of Ethereum heists from 2016–2022. The author also conducted some micro and macro analysis of the money laundering activities on Ethereum. For example, it was disclosed that money launderers often conceal the origins of stolen funds by swapping tokens on DeFi platforms and increase their anonymity by creating counterfeit tokens to launder money.

Moreover, rule-based methods have also been applied to detect such activities. Oad et al. \cite{oad2021blockchain} proposed blockchain-enabled transaction scanning (BTS) for money laundering detection, which specified rules to detect suspicious patterns. Sun et al. \cite{sun2022lstm} labeled mixing transactions based on rules then presented an LSTM Transaction Tree Classifier solution.

\emph{Summary.} Machine learning models have been effectively used to detect money laundering activities in Bitcoin transactions. Tree-based models and well-designed GNN have shown good performance. While most studies used dataset provided by the industry, we believe that crawling transactions or accounts related to mixing services to build updated dataset and investigating money laundering in other DeFi components will also be beneficial. 

\subsection{Phishing Scam} 
\label{sec:phishing}
\subsubsection{Patterns of Phishing Scam}
Phishing scams have been a long threat in traditional online environments, involving the deception of users to reveal sensitive information like private keys or passwords \cite{dhamija2006phishing}. These scams often involve the creation of websites, emails, or messages that closely mimic legitimate platforms. The rise of DeFi has led to the evolution of phishing scams, introducing new variations to exploit unsuspecting victims \cite{wu2020phishers}. In these updated schemes, victims are enticed to transfer cryptocurrencies directly to a malicious address. Meanwhile, traditional phishing techniques, through emails and websites, continue to be employed as well.

In addition, DeFi ecosystem has seen the emergence of a similar type of scam known as the giveaway scam \cite{vakilinia2022cryptocurrency}. Scammers entice viewers during live streams on platforms such as YouTube by promising to return a higher percentage of their contributions to a specific cryptocurrency wallet address, but they do not keep these promises eventually.

\subsubsection{Detection of Phishing Scam}
Various advanced techniques has been used to detect phishing websites and addresses.

(1) {\it Detection of Phishing Websites and Giveaway Scams - } Researchers have employed various methods to detect phishing websites and giveaway scams in DeFi. Phillips and Wilder \cite{phillips2020tracing} used DBSCAN clustering \cite{deng2020dbscan} to identify advance-fee and phishing scams on websites, analyzing public social media and blockchain data to gain insights into the behaviors of scammers and victims. Vakilinia et al. \cite{vakilinia2022cryptocurrency} investigated cryptocurrency giveaway scams on YouTube live streams, identifying suspicious domain names and corresponding Bitcoin and Ethereum addresses, and discussed countermeasures. Li et al. \cite{li2023double} developed CryptoScamTracker, a giveaway scam detection system using Certificate Transparency Logs to identify potential scams. Recently, Li et al. \cite{li2023understanding} utilized the AllenNLP model \cite{gardner2018allennlp} to analyze Twitter contents and detect Twitter giveaway scam lists. They developed CryptoScamHunter, an automated scam detection system that identified giveaway scams and extracted associated cryptocurrency addresses on Bitcoin, Ethereum, Binance, Cardano, and Ripple. Another recent research \cite{saha2023demystifying} analyzed 823 Twitter accounts promoting deceptive NFT collections through giveaways, finding 36\% to be fraudulent. They used Decision Tree, LR, SVM, and RF to detect phishing URLs shared by fraudulent NFT collections, identifying 382 new fraudulent NFT projects on Twitter. However, there is a lack of machine learning techniques that utilize on-chain information to detect these scams.

(2) {\it Detection of Phishing Addresses - } Considering the new form of phishing in DeFi, scholars have shifted to directly detecting phishing addresses. 

First, various features are extracted from the transaction graphs. Chen et al. \cite{chen2020phishing} converted Ethereum transaction networks into graphs, extracting statistical features from nodes and their neighbors, and employed the lightGBM-based Dual-sampling Ensemble algorithm for direct phishing address detection. Recently, Wan et al. \cite{wan2023early} extracted features from the local network and transaction-related time series data, encompassing transaction amount and timestamp. These features enabled a early-stage detection of Ethereum phishing addresses. Ghosh et al. \cite{ghosh2023investigating} investigated the significance of transactional, structural, and temporal features in Ethereum phishing detection. Their study provided recommendation to better exploit structural and temporal features in future. 

After that, node embedding methods, such as Node2Vec \cite{grover2016node2vec} and graph embedding methods, such as Graph2Vec \cite{narayanan2017graph2vec} are used to extract richer features from the transaction graphs \cite{yuan2020detecting, wu2020phishers, xia2022phishing, yuan2020phishing, wang2021tsgn, yu2023streaming}. These features were feed into some tree-based models, such as RF and LightGBM as classifiers ultimately. Later, researchers \cite{wen2022hide, wen2021transaction} introduced phishing hiding frameworks to measure general detection model robustness. 

Moreover, researchers have made significant progress in detecting phishing addresses using GNN and other deep learning methods. Multi-channel GNN \cite{zhang2021blockchain}, heterogeneous GNN \cite{kanezashi2022ethereum}, EGAT \cite{zhou2023detecting} and more specially designed modules \cite{shen2021identity, zhou2022behavior, li2022ttagn, yu2022mp, fu2022ct, zhou2023detecting, tang2022semi} have been developed to improve phishing detection. These methods incorporate comprehensive information, including transaction, temporal, and structural data, into GNN to enhance model capability. Recently, Hu et al. \cite{hu2023bert4eth} proposed BERT4ETH, a pre-trained transformer capable of capturing dynamic sequential patterns within Ethereum transactions, enhancing phishing account detection and de-anonymization tasks.

\begin{figure}[!htb]
\centering 
\setlength{\abovecaptionskip}{0cm}
\setlength{\belowcaptionskip}{0cm}
\includegraphics[scale=0.53]{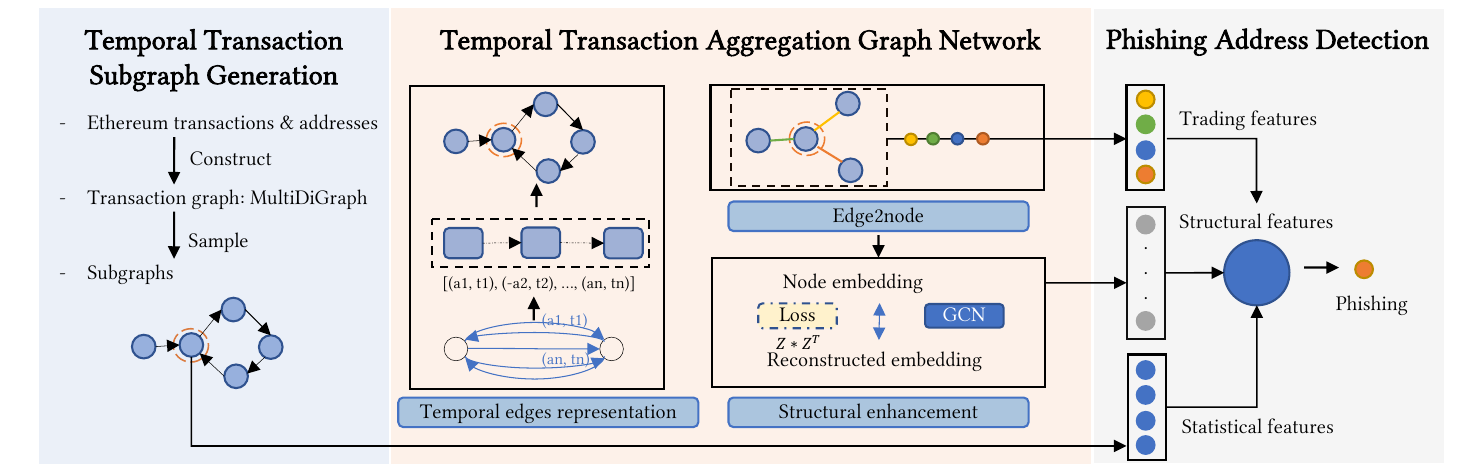} 
\caption{Process of Detecting Phishing Scams (Adapted from \cite{li2022ttagn}).} 
\label{fig:phishing} 
\vspace{-1em}
\end{figure}

As an illustrative example, we explore the state-of-the-art work, TTANG \cite{li2022ttagn} for phishing detection using graph-related techniques. \autoref{fig:phishing} shows the overall architecture of TTANG. It comprises three modules: transaction graph generation, network embedding learning, and phishing address detection. In the first module, Multiple Edges Directed Graphs (MultiDiGraphs) are constructed to model large-scale Ethereum transactions, with a random walk strategy for sampling. The network embedding is learned through three objectives: temporal edges representation using LSTM to capture temporal transaction patterns, edge2node module to aggregate edge representations around each node using Attention, and structural enhancement through combining edge2node features with statistical features and using GCN for learning comprehensive node representation. A LightGBM model is applied as the classifier for Ethereum phishing addresses classification. For evaluation, the authors crawled "phishing" labels from Etherscan before July 2021, and three datasets with sizes of 30k, 40k and 50k nodes were sampled. TTANG achieved good performance on all of them. 

\emph{Summary.} Scholars have designed various techniques for detecting phishing addresses, but the detection of giveaway scams remains relatively limited. Graph-related techniques offer advantages in capturing comprehensive structural information for detecting phishing addresses. Specifically, embedding methods based on random walks and multiple GNN approaches have shown good detection performance. Again, there is potential for more advanced models to be developed specifically to address phishing targeting other than Ethereum, such as NFTs and other DeFi components.

\subsection{Pump and Dump}
Pump and dump (P\&D) schemes artificially inflate asset prices through coordinated buying and spreading misleading information \cite{kamps2018moon}. In DeFi, researchers have studied P\&D schemes since 2018, examining influencing factors, impact, and underlying mechanisms \cite{hamrick2018economics, hamrick2021examination, li2021cryptocurrency, dhawan2021new}. A few P\&D schemes resemble wash trading and rug pulls by occurring immediately after the project's launch, while scammers also target small-cap coins that have been in the market longer. These coins allow them to exert greater control over price movements and maximize their profits \cite{xu2019anatomy}. In addition to pump and dump schemes occurring within a project's lifecycle, there are instances of re-branding or the emergence of new or upgraded projects promising a "better future." For example, the liquidated Three Arrow Capital attempted to launch a new exchange named OPNX\footnote{\url{https://cryptonews.com/news/three-arrows-capital-founders-shake-up-crypto-industry-with-new-crypto-exchange-after-acquiring-coinflex-assets-heres-what-you-need-know.htm}. Accessed on Aug 24, 2023}. 

\subsubsection{Processes of P\&D}
Researchers have found that on-chain P\&D schemes are often closely linked to activities on social media platforms \cite{xu2019anatomy, victor2019cryptocurrency, la2020pump, la2023doge}. A typical process of P\&D in DeFi involves several stages as below: 

\begin{enumerate}
\item The P\&D organizer recruits members through public groups on social media, such as Telegram or Discord and forums, such as Bitcointalk and Reddit.

\item Acting as the group's admin, the admin pre-announces the pump details, including the exact time and date of the pump, along with the rules that participants must adhere to, however, keeping the target coin undisclosed.

\item \textit{Pump:} The admin reveals the target coin as an OCR-proof image and urges members to rapidly buy and hold. This results in a rapid surge in its price within seconds or minutes.

\item \textit{Dump:} As the price falls, panic selling is triggered while the admin may still encourage buying. Although there may be a temporarily boosting of the price, it typically reverts back to its original level or even lower.

\item Within half an hour of the P\&D, the admin may post a review on the coin's price change and selective information. Only details that support this illusion are provided, perpetuating the P\&D schemes.

\end{enumerate}
The processes of P\&D schemes involve coordinated efforts to organize social media channels and manipulate the price of an asset. By gaining an understanding of these mechanisms, researchers can develop strategies to detect them.

\subsubsection{Detection and Prediction of P\&D}
While previous analyses provide insights into P\&D schemes from economic and process, AI techniques have also been used to detect and predict P\&D target coins, events, and user groups.

(1) {\it Detecting P\&D target coins -} Detecting the target coins of a P\&D scheme is crucial for preventing fraudulent activities. Xu and Livshits \cite{xu2019anatomy} conducted an empirical study on P\&D activities in cryptocurrency markets. The authors investigated 412 P\&D events in the cryptocurrency space occurring between June 2018 and February 2019 across multiple crypto exchanges. By tracking Telegram channels associated with P\&D activities, they analyzed the corresponding crypto market movements. Additionally, they targeted at predicting the likelihood of a pump for all coins listed on an exchange before a pump occurs. To achieve this, they proposed various features, such as market cap, returns, volumes, and existence time of the coin. RF and Generalized Linear Models were developed as predictors, with RF achieving a better performance.

Similarly, Nghiem et al. \cite{nghiem2021detecting} addressed the temporal continuity of pumps occurring within a short duration and utilized neural network-based architectures incorporating market and social media signals for accurate predictions. Hu et al. \cite{hu2023sequence} further enhanced the prediction by introducing a sequence-based neural network that leverages a positional attention mechanism to encode pump history of channels.

(2) {\it Detecting P\&D events -} Scholars have also employed sophisticated statistical and machine learning techniques to detect and prevent P\&D activities in DeFi. Kamps and Kleinberg \cite{kamps2018moon} introduced a framework for P\&D detection with a thresholding algorithm to identify suspicious points of anomalous trading activity in exchange. La Morgia et al. \cite{la2020pump} extended this work by applying machine learning models, including RF and LR, to detect P\&D events using ground truth extracted from social media platforms. They later proposed real-time methods using RF and AdaBoost \cite{la2023doge}. Victor and Hagemann \cite{victor2019cryptocurrency} detected potential fraudulent events by concentrating on recurrently announced pumps while filtering out short-term signals and investment advice. They achieved strong results using XGBoost.

(3) {\it Detecting P\&D user groups -} Furthermore, scholars developed methods to detect the user groups involved in P\&D schemes. Chen et al. \cite{chen2019detecting} developed an enhanced apriori algorithm specifically designed for identifying user groups engaged in P\&D activities. Through their investigation of the transaction history of Mt. Gox, they identified numerous abnormal trading records with suspicious behaviors and unusual trading prices.

\emph{Summary.} First, data collection from social media platforms is crucial for identifying P\&D schemes in DeFi, providing clear and traceable signs of P\&D occurrences. Second, statistical and machine learning methods have showcased good detection capabilities for identifying P\&D events and predicting targeted coins. However, AI methods for identifying P\&D user groups are lacking, leaving room for future research in this area.  

\subsection{Summary of Maturity \& Decline Stage Fraud Detection}
Through the review of fraud detection in the project maturity and decline stage, we have the following findings.
\begin{itemize}
    \item Temporal transaction data is now crucial due to the evolving DeFi ecosystem and increased volumes, leading to a dynamic environment and new fraud challenges. To address this, a few sequence-based models have arisen as adopted techniques \cite{sun2022lstm, li2022ttagn, hu2023sequence}. Further approaches that leverage more sequence-based models and time series statistical analysis are needed to enhance fraudulent behavior identification.
    
    \item Deep learning methods, particularly graph-based models, are extensively employed when ample transaction data is available. Unlike earlier stages, advanced graph-related techniques, including node embedding, graph embedding, and well-designed GNN models, have seen growing use to detect fraud in DeFi markets. Heterogeneous graphs and associated learning methods have been for graphs with various node types like wallets and smart contracts. However, the rapid market changes and transaction volume pose a challenge in selecting suitable methods for filtering subgraphs from complete transaction graphs. Addressing this challenge necessitates research to develop efficient and scalable approaches for processing complex and dynamic graph structures.
\end{itemize}

In general, fraud detection has received more attention in this stage than in previous stages, leading to the development of various advanced methods. However, we recognize that the primary focus of detection is on major cryptocurrencies such as Bitcoin and Ethereum. This limitation could be due to ground truth data constraints or the fact that only a few other DeFi components have reached this stage. More detection techniques for other components, such as some NFT collections, are expected to be developed.

\section{Discussion}
\label{sec:discussion}
In this section, we summarize and discuss the characteristics of DeFi frauds, their comparisons with CeFi, and the state-of-the-art detection methods.

\subsection{Organizers and Purposes of DeFi Frauds}
\begin{figure}
\centering 
\includegraphics[scale=0.58]{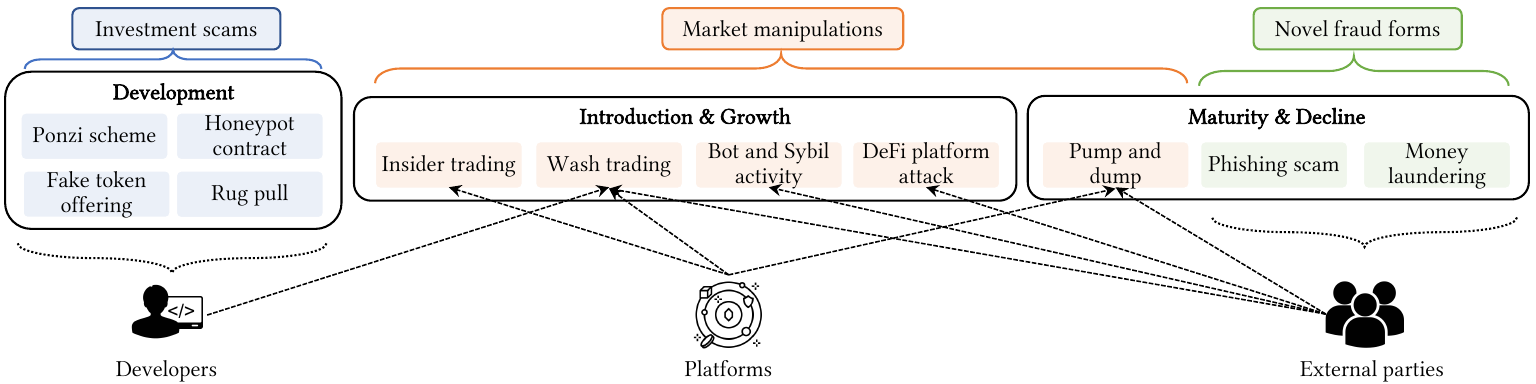} 
\caption{Organizers and Purposes of DeFi Frauds.} 
\label{fig:organizer} 
\end{figure}

When study the DeFi frauds, we notice some correlations between the fraud organizers, the purpose, and the occurrence stages. \autoref{fig:organizer} illustrates their correlations.

Organizers of frauds within the DeFi ecosystem can be categorized into three distinct groups: project developers, DeFi platforms, and external actors. Throughout different stages of DeFi projects, each group exhibits a unique intent to perpetrate fraudulent activities. First, in the development stage, the primary organizers of frauds are the project developers themselves in this stage. They either create fraudulent projects outright or hide fraudulent logic into smart contract codes, with the intent of designing scams from the start. Second, during the introduction and growth stages, multiple players can organize frauds in this phase. Developers engage in wash trading to increase trading volume, raise prices, and attract new investors. The design of some exchanges may also encourage participants to engage in wash trading. Certain NFT marketplaces, for example, provide reward systems that encourage participants to engage in wash trading. Meanwhile, exchanges can engage in insider trading by exploiting information obtained prior to the project's launch. External parties can also contribute to frauds in this phase by creating bot and Sybil accounts and launching DeFi platform attacks. Third, during the maturity and decline stages, more external players enter, gaining the ability to commit fraud. Leveraging DeFi's decentralized and anonymous nature, criminals exploit it for engaging in illicit activities and orchestrating fraudulent events using social media. Additionally, external players may collaborate with DeFi platforms to organize specific frauds during this phase.

We also notice the underlying purposes of these frauds are correlated with their stages. First, during the development stage, investment scams are prevalent, which often deceiving investors through false information and promises. Subsequently, as projects are introduced, market manipulation comes to the fore, often targeting early-stage or small-cap DeFi projects due to their susceptibility to manipulative activities. Meanwhile, criminals use high-cap projects like Bitcoin and Ethereum to conduct money laundering and phishing scams. They have opportunities to create new forms of these frauds due to the vulnerabilities in the blockchain system and the unique characteristics of DeFi.

These correlations make it possible to alternative taxonomies of DeFi frauds, such as categorizing frauds by purpose or organizer. While they can be useful in classifying DeFi frauds, there are some overlaps and limitations. As we discussed, certain frauds may be organized by multiple groups with different motives. In comparison, our taxonomy offers a systematic and comprehensive approach to studying and detecting DeFi frauds by taking into account various stages of a project's life cycle. This taxonomy allows regulators to tailor interventions to specific stages, market participants to proactively evaluate projects and avoid frauds, and researchers to gain insights into the development of DeFi projects and the current state of fraud detection.

\subsection{Comparison of Fraud Detection: CeFi vs. DeFi}
After studying frauds and their detection methods in DeFi, we compare them to their counterparts in CeFi. These two financial ecosystems have both similarities and differences in terms of fraud types and detection approaches.

As discussed in \Cref{sec:factors}, DeFi and CeFi have many similar and different characteristics. Due to these characteristics, frauds in these two financial ecosystems also exhibit both similarities and differences. First, during the development stage, new financial products go through a rigorous development process that frequently necessitates regulatory approvals and compliance with financial laws and regulations in CeFi \cite{legal_match, legal_sing}. This strict regulatory scrutiny may limit the occurrence of frauds. Conversely, DeFi lacks centralized regulatory oversight, which could lead to vulnerabilities in fraud detection in its early stages. Second, during the growth stage in CeFi, there is a notable increase in adoption and market share. Financial institutions leverage their customer base, reputation, and marketing efforts to attract new customers, with main frauds observed in this stage including bank fraud, corporate fraud, and insurance fraud \cite{al2021financial}. In comparison, DeFi projects have the potential for rapid growth due to their decentralized nature and high user yields. This expansion, however, may introduce new risks and scalability issues, particularly with the use of fast-speed bots \cite{zwang2018detecting}. Third, in the maturity and decline stages, certain types of fraud seen in CeFi are also seen in DeFi, resulting in new deployment forms due to the decentralized and anonymous nature of DeFi.

Furthermore, fraud detection between CeFi and DeFi highlights significant differences and similarities in their techniques. First, in CeFi, data acquisition challenges lead to the common use of statistical models and traditional machine learning algorithms, such as SVM, Naive Bayes, and Random Forest, for fraud detection, while the adoption of deep learning models may be less prevalent \cite{al2021financial}. Meanwhile, from the data privacy perspective, scholars may also face the challenge to acquire enough data. Conversely, DeFi's rich on-chain data enables a diverse range of fraud detection models, including various machine learning and graph-based algorithms, effectively identifying fraudulent activities. Second, a common trend in both fields is the growing use of unsupervised and semi-supervised models, which reflects the scarcity of labeled data in both CeFi and DeFi ecosystems \cite{hilal2022financial}. Because of the scarcity of labeled data, they must rely on these alternative methods for effective analysis. Third, NLP methods are widely utilized in both CeFi and DeFi. In CeFi, advanced language models like FinBERT \cite{araci2019finbert} are employed for sentiment analysis and extracting information from financial texts, such as financial news and financial statements. Meanwhile, in DeFi, besides text analysis on DeFi project documents, NLP methods have been applied to analyze contract codes, enabling the understanding and detection of fraudulent contracts.

By examining these various aspects of fraud and their detection methods in CeFi and DeFi, we gain insights into the similarities and differences between these two fields. This enables us to better address fraud risks and develop robust strategies for a secure and trustworthy financial ecosystem.

\subsection{Summary of Current State of DeFi Fraud Detection}
In this subsection, we provide a summary of the state-of-the-art DeFi detection methods, in order to understand the current status of fraud detection within the DeFi domain. \autoref{tab:state-of-the-art-summary} outlines state-of-the-art fraud detection techniques, encompassing the detection task, selected paper, utilized dataset, features employed, detection methods, and achieved performance. The dataset information includes the positive: negative sample ratio or a description if clarity is lacking. In cases where AI-powered detection methods are absent, key detection outcomes are presented through manual extraction or statistical estimation.

\begin{table}[htbp]
\centering
\setlength{\abovecaptionskip}{0cm}
\setlength{\belowcaptionskip}{-0.2cm}
\caption{Summary of State-of-the-art DeFi Frauds Detection Methods.}
\label{tab:state-of-the-art-summary}
\resizebox{\textwidth}{!}{%
\begin{tabular}{m{2.5cm}<{\centering}m{2.8cm}<{\centering}m{1.7cm}<{\centering}m{4.3cm}<{\centering}m{3.5cm}<{\centering}m{2.4cm}<{\centering}m{3.4cm}<{\centering}}
\hline
\textbf{Fraud Type} & \textbf{Task}& \textbf{Paper} & \textbf{Dataset} & \textbf{Features} & \textbf{Detection Methods} & \textbf{Performance} \\
\hline
\multirow{3}{*}{Ponzi scheme} & Detect Bitcoin High Yield Investment Programs (HYIPs) & Bartoletti et al. (2018) \cite{Bartoletti_2018} & Pos:Neg = 32:6,400 & Lifetime, activity days, number of incoming/outgoing transactions, etc.  & RIPPER, Bayes Network, Random Forest & Random Forest detected 18 Ponzi schemes out of 20, producing 81 false positives \\
\cmidrule{2-7} 
&  Detect Ethereum Ponzi contracts & Zheng et al. (2022) \cite{Zheng_2022} & Pos:Neg = 314:6,184 & Bytecode, semantic, developers & Multi-view Cascade Ensemble model & 0.951 Precision, 0.674 Recall, 0.789 F1-score \\
\hline
Honeypot contract & Detect Ethereum honeypot contracts & Chen et al. (2020) \cite{chen2020honeypot} & Pos:Neg = 616:218,250  & N-gram from opcodes & LightGBM & 0.993 Precision, 0.890 Recall, 0.939 F1-score, 0.993 AUC (with undersampling Pos:Neg=1:10) \\
\hline
Fake token offering & Detect scam ICO projects & Bian et al. (2018) \cite{bian2018icorating} & 1,482 projects & White papers, funding teams, websites, and GitHub repositories & Neural network & 0.83 Precision, 0.77 Recall and 0.80 F1-score (with 47\% positive data points) \\
\hline
\multirow{10}{*}{Rug pull} & Detect scam tokens and liquidity pools & Nguyen et al. (2023) \cite{nguyen2023rug}& Pos:Neg=23,871:1,830 & Token, event, pool, creator, average block-dependent features & XGBoost & 0.982 Precision non-malicious 0.999 Recall malicious 0.990 F1-score \\
\cmidrule{2-7} 
 & Detect NFT rug pulls & Huang et al. (2023) \cite{huang2023deep} & Pos:Neg = 7,487: 2,933 & Time-series data, token transfer logs, and secondary market trades & LR, SVM, and RF
& LR, SVM achieved Precision, Recall, and F1 score >= 0.91 (in 4 days before the rug pulls) \\
\cmidrule{2-7} 
 & Detect Fake Exchange & Xia et al. (2020) \cite{xia2020characterizing} &  Data of 70 popular exchanges & Domain features, reported engine, trade volume, etc. & Leveraging existing reports and using typosquatting generation techniques & Detected more than 1,500 scam domains and over 300 fake apps \\
\hline
Insider trading & Detect insider trading activities & Felez-Vinas (2022) \cite{felez2022insider} & Exchange announcements and transaction data during Sep 2018 to May 2022 & Price movements, returns, price runups & Statistical & Estimated insider trading occurs in approximately 10-25\% of cryptocurrency listings \\
\hline
\multirow{6}{*}{Bot \& Sybil account} & Detect arbitrage bot scams & Li et al. (2023) \cite{li2023understanding} & Pos:Neg = 300:600 & YouTube links & AllenNLP & Precision \& Recall \& F1-score \& Accuracy > 0.90 (on both balanced and imbalanced dataset) \\
\cmidrule{2-7} 
& Detect Sybil accounts & Kumar et al. (2020) \cite{kumar2020edarkfind} & 3 real-life Darknet market dataset, namely Dream Market, Tochka, and Wall Street & Stylometric and content-based features, location and substance information &  Unsupervised multi-view learning & 0.98 Precision \& Recall \& F1-score \& Accuracy\\
\hline
Wash trading & Detect wash trading activities & La Morgia et al. (2022) \cite{la2022nft} & Ethereum ERC-721 NFTs transactions before Jan 2022 & NFT transactions & Heuristics  & Detected wash trading activities on NFT markets with a total volume of \$3 Billion   \\
\hline
DeFi platform attack & Front-running & Varun et al. (2022) \cite{varun2022mitigating}& Ethereum transactions from July 2015 to Nov 2020 \cite{torres2021frontrunner} & Transaction gas price, usage of gas tokens, predicted gas price, etc. & MLP & Accuracy > 0.85 (on all dataset)\\
\hline

\multirow{4}{*}{Money laundering} & Detect suspicious addresses associated with mixing services & Wu et al. (2021) \cite{wu2021detecting} & Labeled address: unlabeled address: 131: 1,635,904 (data in 2014, 2015 and 2016) & Network, account, and transaction-level statistical features & A Positive and unlabeled (PU) learning model (Semi-supervised) & TPR > 0.91, FPR < 0.04 (on all dataset)\\
\cmidrule(lr){2-7}
& Detect money laundering transactions &  Lo et al. (2023) \cite{lo2023inspection} & Elliptic dataset \cite{weber2019anti} & Node embedding, local and aggregated features & Self-supervised GNN node embeddings & 0.972 Precision, 0.721 Recall, 0.828 F1-score, 0.916 AUC \\
\hline
\multirow{5}{*}{Phishing scam} & Detect giveaway websites & Li et al. (2023) \cite{li2023understanding} & Pos:Neg = 300:600 &Twitter lists  & AllenNLP & Precision \& Recall \& F1-score \& Accuracy >= 0.99 \\
\cmidrule{2-7}
& Detect phishinig accounts & Li et al. (2022) \cite{li2022ttagn} & Three dataset (D1, D2, D3) with sizes of 30k, 40k and 50k account (Pos:Neg close to 0.3\%) & Accounts, transactions, graph structural & LSTM, edge2node, GCN to create features and LightGBM as classifier &  0.777 Precision, 0.859 Recall, 0.816 F1-score, 0.928 AUC (on D3 dataset) \\
\hline
\multirow{8}{*}{Pump and dump} & Predict targeted coins & Hu et al. (2023) \cite{hu2023sequence} & Pos:Neg = 948:195,665 & Channel, target coin, and market movement signals & Sequence-based neural network (SNN) &  0.260 HR@1, 0.465 HR@5, 0.596 HR@10, 0.797 HR@30 \\
\cmidrule{2-7}
& Detect P\&D events&  La Morgia et al. (2023) \cite{la2023doge} & 317 P\&D happened on Binance & Moving standard deviation of rush orders, number and volume of trades, etc. & Random Forest, AdaBoost &  Random Forest achieved 0.982 Precision, 0.912 Recall, 0.94.5 F1-score in 25 seconds training time \\
\cmidrule{2-7}
& Detect P\&D user groups&  Chen et al. (2019) \cite{chen2019detecting} & Transaction history of Mt. Gox Bitcoin exchange in 2011 to 2013 & Trading actions and times & Data mining &  Detected abnormal trading behaviors and prices \\
\hline
\end{tabular}%
}
\end{table}

First, we find that researchers have utilized various machine learning algorithms, especially tree-based models and graph-based models, to detect various fraud types in DeFi. Notably, Random Forest \cite{rigatti2017random}, XGBoost \cite{chen2015xgboost}, LightGBM \cite{ke2017lightgbm}, and GCN \cite{zhang2019graph} have been widely adopted and have demonstrated their effectiveness. In particular, tree-based models outperformed other models in early-stage DeFi frauds with small data sets, while graph-based models demonstrated effectiveness in later stages with larger dataset and intricate information.

Second, while specific fraud detection tasks have gained considerable attention and demonstrated impressive outcomes, there remain challenges in detecting various fraud types. Notably, the detection of Ethereum Ponzi contracts, money laundering, phishing, and pump and dump has garnered significant research interest, showcasing robust AI-powered detection techniques. However, the detection of other fraud types still presents difficulties. A primary challenge is the limited availability of reliable and comprehensive dataset. For example, detecting money laundering mixer addresses and transactions is constrained by dataset availability, and certain models might rely on outdated data, impacting accuracy and trustworthiness. Additionally, the detection of frauds such as rug pulls on certain DeFi components, insider trading, bot and Sybil accounts, wash trading, and DeFi platform attacks lacks valid machine learning models or is restricted to limited ones. While manual identification and heuristic approaches aid in detecting potential frauds, they possess limitations due to potential human errors and relatively low reliability.

Third, we notice that scholars have explored the potential of self-supervised, semi-supervised, and unsupervised models to overcome data limitations, enabling effective fraud detection with limited data. These advanced learning techniques show promise in mitigating the data scarcity issue and improving the performance of detection models. In addition, another critical challenge is dealing with imbalanced datasets. While certain techniques perform well on balanced dataset, it remains critical for researchers to broaden their investigation and propose effective approaches to the challenges posed by imbalanced data. 

In conclusion, the DeFi fraud detection field shows promising progress, with researchers employing a variety of AI techniques to combat various types of fraud. However, data scarcity remains a major challenge, particularly for certain types of fraud. Future research should focus on addressing the issue of data scarcity and imbalances, as well as employing advanced methods such as self-supervised and unsupervised approaches to improve detection and boost trust in these methods. Further details and additional open issues will be covered in the upcoming section.

\section{Future Directions and Open Issues}
\label{sec:future}
In this section, we delve into potential future directions and open issues in DeFi fraud detection.

\subsection{Advanced AI Detection}
In addition to exploring the fraud landscape, it's also essential to advance AI techniques for DeFi fraud detection.

\textbf{Pre-trained Models and Tranfer Learning}
With limited data in the early stages of the DeFi project, one possible future direction is to leverage the use of pre-trained models. Large-scale pre-trained models like BERT and GPT, known for effectively capturing knowledge from extensive data, have recently achieved remarkable success \cite{han2021pre}. One notable work we have discussed in \Cref{sec:phishing} is the BERT4ETH \cite{hu2023bert4eth}. Scholars can apply more pre-trained techniques for fraud detection in DeFi for improvement of detection performance.

Another related direction is the transfer learning \cite{zhuang2020comprehensive}, which focuses on transferring knowledge across domains in new learning situations. For instance, in CeFi, Lebichot et al. \cite{lebichot2020deep} employed deep transfer learning to detect credit card fraud, transferring classification models learned on e-commerce transactions to face-to-face transactions. In DeFi fraud detection, researchers could consider transferring models from CeFi fraud detection or different DeFi markets, or adapting methods from one fraud type to another.

\textbf{Leveraging Large Language Models} 
In later stages, Large language models (LLMs) can help advance fraud detection, with more comprehensive on-chain transactions and off-chain social media data. LLMs like GPT-3, GPT-4, and LLaMA have recently been a hot research area \cite{zhao2023survey, kasneci2023chatgpt}. Notably, a recent study has developed BloombergGPT \cite{wu2023bloomberggpt}, an LLM specialized in finance. Similarly, researchers can leverage an LLM specialized in DeFi fraud detection. Additionally, an interesting possibility is extending LLMs to generative agents \cite{park2023generative}, which could simulate market conditions and forecast potential fraud in emerging DeFi markets.

\textbf{Choosing the "Right" Model}
In fraud detection across DeFi project life cycle, tree-based, graph-related, and deep learning models have proven effective in diverse tasks \cite{weber2019anti, mazorra2022not, alarab2020competence}. Some models excel with smaller datasets, while others tackle larger and more complex ones. Aside from data volume, the features themselves may have an impact on model capabilities. As a result, an open question arises: How do we choose the "right" detection model for different fraud cases? What is the indicator or threshold for data size that helps us choose the model?

\textbf{Explainable AI Models} Explainable AI is another promising direction that has gained prominence as a significant concern in the financial industry, contributing to regulatory compliance and customer trust \cite{dovsilovic2018explainable}. Various techniques have been developed to tackle the interpretability challenges of AI models, including LIME \cite{picco1999lime}, Anchors \cite{ribeiro2018anchors}, and GraphLIME \cite{huang2022graphlime}. Scholars can use these methods for DeFi fraud detection or create other novel techniques.

\subsection{Exploring the DeFi Fraud Landscape}
As the DeFi ecosystem continues to evolve, new projects and markets are emerging. It is important to continuously explore the DeFi fraud landscape to maintain the safety of the DeFi system.

\textbf{Emergence of DeFi Frauds} The rapidly evolving DeFi landscape is giving rise to novel fraud types. One example is the issue of tax evasion in DeFi. Scholars believe that cryptocurrencies, which are used as payment tools in many DeFi projects, pose a significant risk of tax evasion \cite{houben2018cryptocurrencies}. However, there are no statistical estimates of the severity of the problem, nor are there effective methods for detecting such fraud. To identify patterns and transactions that may indicate tax evasion, on-chain pattern recognition and AI-powered methods can be used. 

Meanwhile, current research on fraud detection in DeFi has primarily focused on well-established blockchains like Bitcoin \cite{oad2021blockchain, vassallo2021application} and Ethereum \cite{li2022ttagn, chen2021improving}. Only a few studies have considered other blockchains \cite{cernera2022token, zheng2022unravelling}. However, as the DeFi ecosystem evolves, new markets on other blockchains, such as Binance Coin and Solana, are also emerging, thereby necessitating scholars to expand their investigations beyond the major chains. Furthermore, as DeFi matures, it becomes increasingly critical to examine the risks associated with emerging components, such as stablecoins \cite{mita2019stablecoin} and DAOs \cite{hassan2021decentralized}. While governance concerns have been addressed in existing literature \cite{chohan2017decentralized, gu2021empirical}, there exists a need for scholars to delve into real fraud cases on these components.

\textbf{Comprehensive Information Utilization} Although current DeFi fraud detection methods have advanced, there is still room for improvement by tapping into more comprehensive data. First, cross-chain analysis, extending beyond individual blockchains, could uncover hidden patterns across various chains and markets. While cross-market data has been used for detecting Sybil accounts \cite{kumar2020edarkfind}, this approach could be expanded to identify other fraud types like money laundering, to track funds and unveil market correlations. Second, another avenue involves examining special accounts, such as whale accounts. ChainAnalysis reported that 4\% of crypto whales engage in criminal activities, holding a collective \$25 billion\footnote{\url{https://blog.chainalysis.com/reports/2022-crypto-crime-report-preview-criminal-balances-criminal-whales/}. Accessed on Jun 12, 2023.}. To uncover fraud signals linked to such accounts, scholars can analyze both on-chain and off-chain behaviors. Third, while some studies have gathered insights from platforms like Telegram and Twitter \cite{xu2019anatomy, saha2023demystifying}, exploring more diverse sources like Reddit and other online forums holds potential for broader insights.

\textbf{Real-time Analysis}
Real-time analysis is required to combat DeFi fraud proactively. Nonetheless, there is limited existing research that emphasizes real-time detection or the efficiency of detection methods \cite{Zheng_2022, la2023doge}, which more researchers should consider in order to identify and predict potential frauds before they cause significant damage. Meanwhile, there has been little research attention directed toward the early detection of frauds \cite{cheng2023evolve}, which is also worth considering. Scholars can also develop early-warning systems that send alerts to users, exchanges, and regulators.

\subsection{Privacy and Regulation}
The anonymity of blockchain transactions and the absence of Know Your Customer (KYC) procedures present challenges for effective regulation in DeFi. While some DeFi exchanges have begun enforcing KYC processes, most crypto exchanges still lack clear KYC policies\footnote{\url{https://www.coindesk.com/markets/2019/03/27/most-crypto-exchanges-still-dont-have-clear-kyc-policies-report/}. Assessed on Aug 16, 2023.}. However, the imposition of KYC is debated due to the decentralized nature of DeFi. We advocate for thoughtfully designed and implemented KYC policies in DeFi that strike a balance between user privacy and regulatory requirements. Applying zero-knowledge techniques\footnote{\url{https://ethereum.org/en/zero-knowledge-proofs/}. Accessed on Aug 28, 2023.} may bridge the gap here by proving the truth of the information without revealing underlying data. This technology is expected to mature further in the future.

Moreover, with more KYC procedures being incorporated into DeFi, the issue of privacy-preserving AI \cite{shokri2015privacy} emerges as another open challenge for DeFi fraud detection methods. To address this issue, researchers need to find a middle ground between privacy and transparency in DeFi, enabling secure data analysis and decision-making while protecting sensitive user information.

Furthermore, there are other regulatory challenges in DeFi, such as smart contract auditing \cite{he2020smart} and oracle problems \cite{caldarelli2020understanding}. Collaborative efforts involving computer scientists, industry stakeholders, and legal experts are essential to establish appropriate regulatory frameworks in the DeFi ecosystem.

\section{Conclusion}
\label{sec:conclusion}
In conclusion, this survey provides a comprehensive taxonomy of financial frauds in the DeFi ecosystem across different project stages. We have analyzed various fraudulent activities with a focus on on-chain patterns, characteristics and AI-powered detection methods. Our review reveals a strong relationship between DeFi frauds and project stages, encompassing fraud organizers, purposes, and detection methods. To further advance fraud detection in DeFi, future studies should focus on exploring emerging fraud types and new DeFi components as the ecosystem evolves. Leveraging advanced AI-powered techniques, such as pre-trained models and explainable models, can also help detect fraud. By implementing these insights, researchers, practitioners, and regulators can collaborate to establish a more trustworthy DeFi ecosystem.

\bibliographystyle{ACM-Reference-Format}
\bibliography{fraudbib}

\end{document}